\documentclass[]{IEEEtran}
\pdfoutput=1
\usepackage{tikz}
\usepackage{longtable}
\usepackage{supertabular}
\usepackage[sumlimits,intlimits]{amsmath}
\usepackage{amsfonts,amssymb}%
\usepackage{bm}%
\usepackage{graphicx,graphics}%
\usepackage[english]{babel}
\usepackage[noadjust]{cite}%
\usepackage{color}%
\usepackage{url}
\usepackage{verbatim}
\usepackage{balance}
\usepackage{multirow}
\usepackage{multicol}
\usepackage{stfloats}
\usepackage{morefloats}
\usepackage{xspace}
\usepackage{enumerate}
\usepackage{rotating}
\usepackage[tight]{subfigure}
\usepackage{algorithm}
\usepackage{algorithmicx, algpseudocode}
\usepackage{pdfpages,hyperref}
\usepackage[numbers]{natbib}

\usepackage{array}
\newcolumntype{L}[1]{>{\raggedright\let\newline\\\arraybackslash\hspace{0pt}}m{#1}}
\newcolumntype{C}[1]{>{\centering\let\newline\\\arraybackslash\hspace{0pt}}m{#1}}
\newcolumntype{R}[1]{>{\raggedleft\let\newline\\\arraybackslash\hspace{0pt}}m{#1}}

\ifCLASSOPTIONpeerreview

\fi

\begin{document}

\title{Timing and Carrier Synchronization in Wireless Communication Systems: A Survey and Classification of Research in the Last Five Years}

\author{Ali A.~Nasir, Salman~Durrani, Hani Mehrpouyan, Steven D. Blostein, and Rodney~A.~Kennedy%
\thanks{Ali A. Nasir, Salman Durrani and Rodney A. Kennedy are with the Research School of Engineering, Australian National University, Canberra, ACT 2601, Australia (Email: \{ali.nasir, salman.durrani, rodney.kennedy\}@anu.edu.au). Hani Mehrpouyan is with the Department of Electrical and Computer Engineering, Boise State University, Idaho, USA (Email: hani.mehr@ieee.org). Steven D. Blostein is with the Department of Electrical and Computer Engineering, Queen's University, Canada (Email: steven.blostein@queensu.ca).}
}

\maketitle

\vspace{-12pt}

\begin{abstract}
Timing and carrier synchronization is a fundamental requirement for any wireless communication system to work properly. Timing synchronization is the process by which a receiver node determines the correct instants of time at which to sample the incoming signal. Carrier synchronization is the process by which a receiver adapts the frequency and phase of its local carrier oscillator with those of the received signal. In this paper, we survey the literature over the last five years (2010-2014) and present a comprehensive literature review and classification of the recent research progress in achieving timing and carrier synchronization in single-input-single-output (SISO), multiple-input-multiple-output (MIMO), cooperative relaying, and multiuser/multicell interference networks. Considering both single-carrier and multi-carrier communication systems, we survey and categorise the timing and carrier synchronization techniques proposed for the different communication systems focusing on the system model assumptions for synchronization, the synchronization challenges, and the state-of-the-art synchronization solutions and their limitations. Finally, we envision some future research directions.
\end{abstract}



\ifCLASSOPTIONpeerreview
    \newpage
\fi

\section{Introduction}

\textbf{Motivation:} The Wireless World Research Forum (WWRF) prediction of \textit{``seven trillion wireless devices serving seven billion people by 2020"}~\cite{David-2012} sums up the tremendous challenge facing existing wireless cellular networks: intense consumer demand for faster data rates. Major theoretical advances, such as the use of multiple antennas at the transmitter and receiver (MIMO)~\cite{Blostein-2003,Gesbert-2010}, orthogonal frequency-division multiple access (OFDMA)~\cite{Hwang-2009}, and cooperative relaying~\cite{Laneman-04-A,Pabst-2004,Yang-2009} have helped meet some of this demand and have been quickly incorporated into communication standards. These technologies also form a core part of next generation cellular standards, 5G, which is under development~\cite{Andrews-2014,metis-2013}.

In order to fulfill the demand for higher data rates, a critical requirement is the development of accurate and realizable synchronization techniques to enable novel communication paradigms. Such synchronization techniques allow communication systems to deliver higher data rates, e.g., through the use of higher order modulations and utilization of cooperative communication schemes. Hence, there has been considerable research recently in synchronization techniques for novel communication strategies.

\textbf{Aim:} The aim of this paper is to provide a survey and classification of the research in the field of synchronization for wireless communication systems that spans the last five years (2010-2014). This is not an easy task given the large number of papers dealing with synchronization and its associated challenges in both current and emerging wireless communication systems. \textit{The critical need for such a survey is highlighted by the fact that the last comprehensive survey paper on synchronization was published nearly a decade ago~\cite{Morelli-2007}}. While survey papers on synchronization for wireless standardization have recently appeared~\cite{Magee-2010,Tipmongkolsilp-2010,Bladsjo-2013}, these surveys do not overview the state-of-the-art published research.

In this survey, we overview the relationships among the published research in terms of system model and assumptions, synchronization challenges, proposed methods, and their limitations. We also highlight future research directions and important open problems in the field of synchronization. The main intended audience of this survey paper is anyone interested in or already working in synchronization. Our hope is that this survey paper would enable researchers to quickly immerse themselves in the current state-of-the-art in this field. Moreover, by highlighting the important open research issues and challenges, we believe the paper would stimulate further research in this field. Since this paper is not intended to be a tutorial on synchronization, we deliberately avoid presenting mathematical details and instead focus on the big picture.

\textbf{Background and Scope:} Synchronization is a common phenomenon in nature, e.g., the synchronized flashing of fireflies or the synchronous firing of neurons in the human brain~\cite{Barbarossa-2007,Tyrrell-2010}. In wireless communications, timing and carrier synchronization is a fundamental requirement~\cite{Meyr-1998}. In general, a wireless receiver does not have prior knowledge of the physical wireless channel or propagation delay associated with the transmitted signal. Moreover, to keep the cost of the devices low, communication receivers use low cost oscillators which inherently have some drift. In this context:
\begin{enumerate}
\item \textit{Timing synchronization} is the process by which a receiver node determines the correct instants of time at which to sample the incoming signal.

\item \textit{Carrier synchronization} is the process by which a receiver adapts the frequency and phase of its local carrier oscillator with those of the received signal.
\end{enumerate}

For instance, requiring two watches to be time synchronized means that they should both display the same time. However, requiring two watches to be carrier synchronized means that they should tick at the same speed, irrespective of what time they show~\cite{Lombardi-2012}. \textit{Note that channel estimation, which is an inherent requirement for synchronization, is not the main focus of this paper}. For a recent survey and tutorial on channel estimation alone, please see~\cite{Liu-2014}.

Major advances in timing and carrier synchronization such as pilot symbol assisted modulation~\cite{Cavers-1991}, are used in present day cellular networks to achieve carrier accuracy of 50 parts per billion and timing accuracy of 1 $\mu$s ($\pm$500 ns)~\cite{Magee-2010}. The requirement in future wireless networks is towards tighter accuracies, e.g., timing accuracy of 200 ns, to enable location-based services~\cite{Tipmongkolsilp-2010}. Hence, there is a need for new and more accurate timing and carrier estimators. In general, in order to quantify the performance of any proposed estimator, a lower bound on the mean-square estimation error can be derived. The bounds are also helpful in designing efficient training sequences. In addition, for multiple parameters needed, say, for the joint estimation of timing and carrier frequency offsets, these bounds include coupling information between the estimation of these parameters. For example, if the bound suggests very low coupling between the estimation of timing and carrier frequency offsets, this implies that these parameters can be estimated separately without any significant loss in the estimation performance. In particular, there usually exist strong coupling between channel and carrier frequency offset estimation and their joint estimation is helpful to achieve improved estimation accuracy~\cite{Mehr-11-A,Nasir-12-A}.



Although timing and carrier synchronization is necessary for successful communication, it cannot provide a common notion of time across distributed nodes. \textit{Clock synchronization} is the process of achieving and maintaining coordination among independent local clocks to provide a common notion of time across the network. Some wireless networks, such as worldwide interoperability for microwave access (WiMAX), are synchronized to the global positioning system (GPS)~\cite{Tipmongkolsilp-2010}. Others, e.g., Bluetooth, wireless fidelity (WiFi), and Zigbee rely on a beacon strategy, where all nodes in the network follow the same time reference given by a master node broadcasting a reference signal~\cite{Tipmongkolsilp-2010}. \textit{In the literature, clock synchronization is considered separately from timing and carrier synchronization and is excluded from this survey}. For recent surveys on clock synchronization, please see~\cite{Djenouri-2015,Serpedin-B-2011,Wu-2011,Freris-2010,Simeone-2008}.

In the literature, timing and carrier synchronization techniques are sometimes considered in conjunction with radio frequency (RF) front-end impairments. \textit{RF impairments} arise as a result of the intrinsic imperfections in many different hardware components that comprise the RF transceiver front-ends, e.g., amplifiers, converters, mixers, filters, and oscillators. The three main types of RF impairments are I/Q imbalance, oscillator phase noise, and high power amplifier (HPA) nonlinearities~\cite{Fettweis-2005}. I/Q imbalance refers to the amplitude and phase mismatch between the in-phase (I) and quadrature (Q) signal branches, i.e., the mismatch between the real and imaginary parts of the complex signal. Oscillator phase noise refers to the noise in an oscillator, mainly due to the active devices in the oscillator circuitry, which introduces phase modulated noise, directly affecting the frequency stability of the oscillator~\cite{Poddar-2013}. The HPA nonlinearities refer to the operation of the HPA in its nonlinear region when working at medium and high-power signal levels. The influence of these RF impairments is usually mitigated by suitable compensation algorithms, which can be implemented by analog and digital signal processing. In this paper, the focus is on timing and carrier synchronization and \textit{RF impairments are outside the scope of this paper}. For a detailed discussion of RF impairments, the reader is referred to~\cite{Schenk-2008}. In cases where RF impairments (typically I/Q imbalance or phase noise) are considered in conjunction with timing and carrier synchronization, they are identified separately in the classification.

\textbf{Methodology}: Synchronization is generally considered as a subfield of signal processing. According to Google Scholar, $9$ out of the top $10$ publication avenues in signal processing are IEEE journals~\cite{google}. Hence, we used the IEEEXplore database to search for papers on timing and carrier synchronization. Synchronization in wireless communication systems is an active area of research and there are a very large number of papers on this topic in IEEEXplore. For example, a general search with the words ``timing synchronization" yields close to $19,000$ papers (admittedly not all papers would fit the scope of this survey).

We selected papers (in December 2014) by searching for words ``frequency offset" OR ``frequency offsets" OR ``timing offset" OR ``timing offsets" in IEEEXplore metadata only. In order to focus on the important recent advances, we limited our search to all journal papers published in the last 5 years only, i.e., from 2010-2014. Also, we limited the search to the following conferences: ICC, GLOBECOM, VTC, WCNC, SPAWC, and PIMRC, because it was found that these conferences contained sufficient numbers of papers to address the synchronization topics.

Using these principles, papers that dealt with timing and carrier synchronization were carefully selected for inclusion in this survey paper. A classification of these papers, with respect to the adopted communication system, is presented in Table~\ref{tab:Coop_MC1}. Some papers were found to study the effect of timing and carrier synchronization on the performance of various communication systems, but they did not directly estimate or compensate for these synchronization impairments. These papers are summarized in Table~\ref{tab:Coop_MC2} for the sake of completeness. However, these papers are not discussed in the survey sections below.

\textbf{Abbreviations and Acronyms}: The list of abbreviations and acronyms used in this paper are detailed in Table \ref{tab:acronyms}. In the paper, in Tables~\ref{tab:SISO_SC}-\ref{tab:60G}, `CSI Req.' column indicates (using Yes/No value) whether channel state information (CSI) is required for synchronization procedure or not ,`CE' column indicates (using Yes/No value) whether algorithm considers channel estimation (CE) or not, `Est/Comp' column indicates whether algorithm only considers estimation (Est) of parameters or also uses the estimated parameters for compensating (Comp) their effect on system bit-error-rate (BER) performance, `N/A' stands for \emph{not applicable}, and `Bound' column indicates whether the paper derives or provides lower bound on the estimation performance.

\textbf{Organization}: The survey is organized as  follows. The selected papers are classified into five categories: (i) single input single output (SISO) (Section~\ref{sec:siso}), (ii) multiple input multiple output (MIMO) (Section~\ref{sec:mimo}), (iii) cooperative relaying (Section~\ref{sec:cooperative}), (iv) multicell/multiuser (Section~\ref{sec:interference}), and (v) other (ultra wide band (UWB) and spread spectrum) communication networks (Section~\ref{sec:other}). \textit{Each category is split into single carrier and multi-carrier (e.g., OFDM) systems}. For each category, we discuss the system model for synchronization, the synchronization challenges and the state-of-the-art synchronization solutions and their limitations. Future research directions and important open problems are highlighted in Section \ref{sec:future}. Finally, Section \ref{sec:con} concludes this survey.
\begin{table*}[p]
\caption{List of common acronyms and abbreviations} \centering
\begin{tabular}{|l|l|}\hline
Acronym & Definition \\ \hline
AF & amplify-and-forward\\
AFD-DFE & adaptive frequency domain decision feedback equalizer \\
AOD & angle of departure \\
BER & bit-error-rate \\
CE & channel estimation \\
Comp & compensation \\
CFO & carrier frequency offset\\
CP & cyclic prefix \\
CSI & channel state information \\
DoA & direction of arrival \\
DF & decode-and-forward\\
DL & direct link \\
DLC-SFC & distributed linear convolutional space frequency code \\
DLC-STC & distributed linear convolutional space time coding \\
DSFBC & distributed space frequency block coding \\
DSTBC & distributed space time block coding \\
DSTC & distributed space time coding \\
Est & estimation \\
FBMC & filter bank multi-carrier \\
FDE & frequency domain equalization \\
FDMA & frequency division multiple access \\
FD-S$^3$ & frequency domain-spread spectrum system \\
FFT & fast Fourier transform \\
Freq. flat & frequency flat \\
Freq. sel. & frequency selective \\
GD-S$^3$ & Gabor division-spread spectrum system \\
HetNet & heterogeneous network \\
IFFT & inverse fast Fourier transform \\
IFO & integer frequency offset \\
IQ & in-phase quadrature-phase \\
IR & impulse radio \\
MAI & multiple access interference \\
MB-OFDM & multiband-OFDM \\
MC & multi carrier \\
MCFOs & multiple carrier frequency offsets \\
MISO & multiple-input-single-output \\
MTOs & multiple timing offsets\\
N/A & not applicable \\
OSTBC & orthogonal space time block coding \\
OWRN & one-way relaying network\\
PHN  & phase noise \\
PUs  & primary users \\
req. & required \\
Rx & receiver \\
SC & single carrier \\
SCO & sampling clock offset \\
SDR & software defined radio \\
SFBC & space frequency block coding \\
SFCC & space frequency convolution coding \\
SFO & sampling frequency offset \\
SIMO & single-input multiple-output \\
STC & space time coding \\
TD-LTE & time division long term evolution \\
TH & time hopping \\
TR-STBC & time reversal space time block code \\
TO & timing offset \\
TS & training sequence \\
TWR & two way ranging \\
TWRN & two-way relaying network\\
Tx & transmitter \\
UFMC & universal filtered multi-carrier \\
WSN & wireless sensor network \\ \hline
\end{tabular}
\label{tab:acronyms}
\end{table*}


\begin{table*}[p]
\vspace{1cm}
\caption{Classification of papers on timing or carrier synchronization, 2010-2014} \centering
\begin{tabular}{|C{2.5cm}|C{2.5cm}|C{5.5cm}|C{5.5cm}|} \hline
\multicolumn{2}{|c|}{{Communication System}} & Single-carrier & Multi-carrier  \\ \hline
\multicolumn{2}{ |c| }{SISO} &
\cite{Lin-14-A},\cite{Shaw-13-A},\cite{Lin-13-A},\cite{Wu-11-A},\cite{Yin-14-A},\cite{Rabiei-11-A},\cite{Colonnese-10-A},\cite{Bai-10-A},
\cite{Colonnese-12-A},\cite{Oh-13-A},\cite{Lin-10-A},\cite{Man-13-A},\cite{Pedrosa-10-A},\cite{Pan-12-A},\cite{Sahinoglu-11-A},\cite{Huh-10-A},
\cite{Man-13-M-A},\cite{Silva-14-A},\cite{Kim-10-Aug-A},\cite{Lin-13-May-A},\cite{Hosseini-13-A},\cite{Zhang-14-Feb-A},\cite{Wu-11-Jul-A},
\cite{Gong-12-A},\cite{Zhang-10-Jul-A},\cite{Hosseini-13-Dec-A},\cite{Nasir-12-PT-A},\cite{Kim-13-A},\cite{Ramakrishnan-12-A},\cite{Dobre-12-A},
\cite{Punchihewa-10-A},\cite{Nasir-11-May-P},\cite{Elgenedy-13-Jun-P},\cite{Yan-11-Jun-P},\cite{Yan-10-May-P},\cite{Obara-10-May-P},\cite{Bhatti-11-Sep-P},
\cite{Zhao-13-Jun-P},\cite{Chen-10-Apr-P},\cite{Popp-13-Sep-P},\cite{Bao-12-May-P},\cite{Wang-10-May-P},\cite{Inserra-10-Sep-P},\cite{Peng-14-Art}
 &
\cite{Li-12-A},\cite{Long-11-A},\cite{Liu-10-A},\cite{Cai-10-A},\cite{Cai-11-A},\cite{Chin-14-A},\cite{Punchihewa-11-A},\cite{Shim-10-A},
\cite{Bai-12-A},\cite{Bassiouni-13-A},\cite{Xie-12-A},\cite{You-10-A},\cite{Dweik-10-A},\cite{Yang-12-A},\cite{Rotoloni-12-A},\cite{Liu-11-A},
\cite{Yuan-14-A},\cite{You-10-O-A},\cite{Zhang-13-A},\cite{Morelli-13-A},\cite{Chen-12-A},\cite{Welden-12-A},\cite{Lmai-14-A}, \cite{Salim-14-A},\cite{Bayon-10-A},\cite{Simon-12-A},\cite{Lee-13-A},\cite{Lim-13-A},\cite{Xu-12-A},
\cite{Morelli-12-A},\cite{Yang-14-A},\cite{Tanhaei-11-A},\cite{Choi-10-A},\cite{Dweik-10-N-A},\cite{Lijun-10-A},\cite{Cvetkovic-13-A},\cite{Rahimi-14-A},
\cite{Jeon-11-A},\cite{Bai-13-A},\cite{Su-13-A},\cite{Wang-12-A},\cite{Oh-11-A},\cite{Chu-14-A},
\cite{Xu-10-A-A},\cite{Tseng-13-A},\cite{Wu-10-A},\cite{Nguyen-10-A},\cite{Morelli-10-A-A},\cite{Carvajal-13-A},\cite{Jing-14-A},\cite{Wang-14-A}, \cite{Lanlan-10-A},\cite{Lu-12-A},\cite{Lee-11-A},\cite{Zhang-12-F-A},\cite{Sheng-10-A},\cite{Ziabari-11-A},\cite{Kim-11-Jun-A},\cite{Wang-10-Mar-A},
\cite{Dweik-12-A},\cite{Peng-13-A},\cite{Wang-13-A},\cite{Marchetti-14-A},\cite{Morelli-14-Apr-A},\cite{Liu-10-May-A},\cite{Li-12-Dec-A},
\cite{Lifeng-10-A},\cite{Kim-10-Jan-A},\cite{He-11-A},\cite{Zhang-12-Feb-A},\cite{Zhu-13-FQ-A},\cite{Xu-13-Jun-A},\cite{Chin-14-May-A},
\cite{Ketseoglou-10-A},\cite{Oliver-12-A},\cite{Sanguinetti-10-A},\cite{Dwivedi-12-A},\cite{Shi-10-A},\cite{Zhang-12-May-A},\cite{Ma-12-A},\cite{Mohebbi-14-A}
\cite{Xu-13-Dec-A},\cite{Kang-10-A},\cite{Shi-10-Nov-A},\cite{Kung-12-A},\cite{Morelli-13-Feb-A},\cite{Viemann-10-A},\cite{Lottici-10-A},\cite{Punchihewa-10-A},
\cite{Morelli-10-Dec-P},\cite{Kume-12-Jun-P},\cite{Li-10-May-P},\cite{Yeh-12-Dec-P},\cite{Smida-11-Dec-P},\cite{Dainelli-11-Jun-P},\cite{Wen-11-Dec-P},
\cite{Lin-13-Jun-P},\cite{Bellanger-12-Dec-P},\cite{Xu-10-Dec-P},\cite{Li-13-Jun-P},\cite{Shahriar-13-Jun-P},\cite{Wang-11-Jun-P},\cite{Yang-12-Dec-P}, \cite{Udupa-13-Jun-P},\cite{Zhang-10-May-P},\cite{Tsai-10-May-P},\cite{Guo-11-May-P},\cite{Gao-10-Sep-P},\cite{Guillaud-13-Apr-P},\cite{Wang-11-Sep-P},
\cite{Hong-11-Sep-P},\cite{Surantha-13-Sep-P},\cite{Andgart-10-Sep-P},\cite{Chang-10-Sep-P},\cite{Chang-12-Sep-P},\cite{Miyashita-12-May-P},\cite{Nasraoui-12-Apr-P},
\cite{Gregorio-10-Apr-P},\cite{Pan-11-May-P},\cite{Nasraoui-11-Sep-P},\cite{Zhang-11-May-P},\cite{Wiegand-12-Sep-P},\cite{Liu-12-Sep-P},\cite{Kiayani-11-Sep-P},
\cite{Thein-13-Sep-P},\cite{Chang-10-May-VTC-P},\cite{Chan-13-Sep-P},\cite{Ishaque-13-Sep-P},\cite{Thein-13-Jun-P},\cite{Liu-14-Nov-Art} \\
\hline
\multicolumn{2}{ |c| }{Multiple Antenna} &
\cite{Jiang-13-A-A},\cite{Zhang-10-N-A},\cite{Zhang-10-May-A},\cite{Du-13-A},\cite{Marey-12-A},\cite{Eldemerdash-13-Dec-P},
\cite{Gao-10-May-P},\cite{Sinha-13-Sep-P},\cite{Yao-11-Mar-P},\cite{Nasir-11-A},\cite{Mohammadkarimi-14-Oct-Art} &
\cite{Salari-11-A},\cite{Choi-14-A},\cite{Kwon-12-A},\cite{Shim-10-D-A},\cite{Younis-10-A},\cite{Liu-10-A-A},\cite{Yu-12-A},\cite{Bannour-12-A}, \cite{Amo-13-A},\cite{Zhang-13-M-A},\cite{Simon-11-A},\cite{Liang-10-A},\cite{Zhang-A-A-14},\cite{Nguyen-11-A},\cite{Kim-11-A}, \cite{Baek-10-A},\cite{Jeon-14-A},\cite{Hsu-12-A},\cite{Jose-13-A},\cite{Chung-10-A},\cite{Baek-11-A},\cite{Choi-11-A},\cite{Narasimhan-10-A},
\cite{Weikert-13-Jun-P},\cite{Jiang-13-Dec-P},\cite{Mahesh-12-Apr-P},\cite{Xu-13-Apr-P},\cite{Lei-11-Sep-P},\cite{Wang-11-Sep-VTC-P},
\cite{Luo-11-Sep-P},\cite{Wang-10-Sep-P},\cite{Jing-14-Sup-Art}\\
\hline
\multirow{6}{*}{Cooperative}
& QF-OWRN &
\cite{Avram-12-Sep-P}&
\\
\cline{2-4}
& AF-OWRN &
\cite{Mehr-11-A},\cite{Nasir-13-A},\cite{Nasir-12-A},\cite{Li-10-A},\cite{Mehrpouyan-11-C-A},\cite{Yadav-10-May-P},\cite{Mehrpouyan-10-Dec-P} &
\cite{Jung-14-A},\cite{Yao-12-A},\cite{Zhang-12-Jun-A},\cite{Jung-14-A},\cite{Zhang-13-Mar-A},\cite{Salim-14-Jun-P},\cite{Zhang-12-May-P},\cite{Won-13-Sep-P}\\
\cline{2-4}
& DF-OWRN  &
\cite{Mehr-11-A},\cite{Nasir-12-A},\cite{Liu-12-A},\cite{Nasir-13-AT},\cite{Wang-11-Feb-A},\cite{Wang-10-Nov-A},\cite{Liu-14-Mar-A},\cite{Nasir-11-Jun-P},
\cite{Wang-12-Jun-P},\cite{Mehrpouyan-10-Dec-P},\cite{Nasir-13-Jun-P} &
\cite{Kim-11-Sep-A},\cite{Huang-10-J-A},\cite{Guo-13-A},\cite{Xiong-13-A},\cite{Huang-10-Nov-A},\cite{Thiagarajan-10-Dec-P},\cite{Rugini-12-Jun-P},
\cite{Wang-12-Dec-P},\cite{Etezadi-10-Dec-P},\cite{Zhang-12-May-P},\cite{Lu-10-Sep-P},\cite{Lu-11-May-P},
\cite{Ponnaluri-10-Dec-P},\cite{Xiao-11-May-P},\cite{Yang-10-Dec-P},\cite{Wang-13-Sep-P},\cite{Zhang-13-Sep-P}\\
\cline{2-4}
& AF-TWRN &
\cite{Chang-12-A},\cite{Jiang-13-A},\cite{Zha-14-A},\cite{Wu-14-Jun-P},\cite{Nasir-14-Jun-SPAWC-P},\cite{Ferrett-12-May-P} &
\cite{Ho-13-A},\cite{Wang-11-A},\cite{Li-13-Dec-P}\\
\cline{2-4}
& DF-TWRN &
\cite{Jain-11-Dec-P} &
\\
\hline
\multirow{11}{*}{Multiuser / Multicell  }
& SC-FDMA uplink &
\multicolumn{2}{c|}{\cite{Darsena-13-A},\cite{Song-11-A},\cite{Meng-10-A},\cite{Kamali-12-A},\cite{Zhu-10-A},\cite{Zhang-Ryu-10-May-A},\cite{Iqbal-14-A},
\cite{Kamali-14-A},\cite{Kamali-11-Mar-A},\cite{Fu-14-A},\cite{Chen-10-May-P}} \\
\cline{2-4}
& OFDMA uplink &
\multicolumn{2}{C{10cm}|}{\cite{Kim-10-A},\cite{Zhang-14-A},\cite{Pengfei-A-10},\cite{Shah-13-A},\cite{Ho-14-A},\cite{Hsieh-11-A},\cite{Zhang-14-M-A},\cite{Zhang-12-A},\cite{Lee-12-A},\cite{Keum-14-A},\cite{Nguyen-14-A},\cite{Estrella-13-A},\cite{Yuejie-11-A},
\cite{Morelli-10-A},\cite{Movahhedian-10-A}\cite{Sourck-11-A},\cite{Song-11-A},\cite{Huang-13-A},\cite{Haring-10-A},\cite{Lin-10-A},\cite{Miao-10-A},
\cite{Bai-12-Jun-A},\cite{Muneer-14-A},\cite{Lee-12-Aug-A},\cite{Peng-12-A},\cite{Fa-13-A},\cite{Hou-12-A},\cite{Sanguinetti-10-A},\cite{Yerramalli-10-Dec-P},\cite{Aziz-11-Dec-P},\cite{Chen-10-Dec-P},\cite{Bertrand-10-May-P},\cite{Chang-10-May-P},
\cite{Wu-10-May-P},\cite{Zhang-10-May-VTC-P},\cite{Xiong-12-Sep-P},\cite{Tu-12-May-P},\cite{Farhang-13-Jun-P}}\\
\cline{2-4}
& CDMA  &
\cite{Xu-13-A} &
\cite{Wang-Coon-11-A},\cite{Bangwon-12-A},\cite{Tadjpour-10-A},\cite{Lin-12-Oct-A},\cite{Manglani-11-A},\cite{Li-10-Dec-P},\cite{Yan-14-Art} \\
\cline{2-4}
& Cognitive Radio &
\cite{Axell-12-A},\cite{Rebeiz-13-A},\cite{Nevat-12-Apr-P}&
\cite{Xu-10-A},\cite{Zeng-13-A},\cite{Chen-12-Jan-A},\cite{Al-Habashna-12-A},\cite{Cheraghi-12-A},\cite{Ding-13-Dec-P},\cite{Zhao-11-Dec-P},
\cite{Zivkovic-11-Dec-P},\cite{Zhou-11-Dec-P},\cite{Liu-12-May-P} \\
\cline{2-4}
& Distributed Multiuser &
&
\cite{Liu-12-Aug-A},\cite{Caus-11-Jun-P},\cite{Zhi-11-May-P},\cite{Sanchez-11-Jun-P}\\
\cline{2-4}
& CoMP &
\cite{Zarikoff-10-A},\cite{Liu-13-Apr-P}  &
\cite{Zhao-14-A},\cite{Liang-12-A},\cite{Iwelski-14-A},\cite{Jiang-14-A},\cite{Tsai-13-A},
\cite{Vakilian-13-Dec-P},\cite{Koivisto-13-Sep-P},\cite{Pec-14-A} \\
\cline{2-4}
& Multicell Interference &
\cite{Coelho-11-Sep-P},\cite{Mochida-12-May-P} &
\cite{Deng-14-A},\cite{Morelli-14-A},\cite{Hung-14-A},\cite{Chang-12-Nov-A},\cite{Yang-13-A},\cite{Liu-13-Apr-WCNC-P} \\
\hline
\multirow{3}{*}{ Other  }
& UWB &
\cite{Erseghe-11-A},\cite{DAmico-13-A},\cite{Erseghe-11-D-A},\cite{Chen-10-May-A},\cite{Oh-10-May-A},\cite{Lv-11-A} &
\cite{You-11-A},\cite{Kim-11-J-A},\cite{Lin-12-A},\cite{Karim-10-A},\cite{Wang-10-A},\cite{Hwang-12-A},\cite{Ye-10-A},\cite{Fan-12-A} \\ \cline{2-4}
& Spread Spectrum &
\cite{Oh-12-A-A},\cite{Benedetto-11-A},\cite{Kohda-13-Sep-P}&
 \cite{Kohd-12-Dec-P},\cite{Jang-13-A} \\
\cline{2-4}
& 60 GHz &
&
\cite{Koschel-12-Sep-P},\cite{Ban-13-Jun-P} \\
\hline
 \end{tabular}
\label{tab:Coop_MC1}

\vspace{1cm}
\caption{Papers on the Effect of Timing or Carrier Synchronization on the System Performance, 2010-2014} \centering
\begin{tabular}{|C{2.5cm}|C{2.5cm}|C{5.5cm}|C{5.5cm}|} \hline
\multicolumn{2}{|c|}{Communication System} & Single-carrier & Multi-carrier  \\ \hline
\multicolumn{2}{ |c| }{SISO} &
\cite{Hassan-11-A},\cite{Tavares-10-A},\cite{Dimitrijevic-12-A},\cite{Li-10-M-A},\cite{Chiu-10-A},\cite{Le-13-A},\cite{Sen-13-A},
\cite{Muller-11-A},\cite{Liavas-12-A},\cite{Benvenuto-10-A},\cite{Hamdi-10-Aug-A},\cite{Kim-10-Dec-A} &
\cite{Huang-10-A},\cite{Hua-13-A},\cite{Han-14-A},\cite{Hamza-14-A},\cite{Hamdi-10-A},\cite{Zheng-10-A},\cite{Mahesh-10-A},\cite{Huo-11-A},
\cite{Martinez-10-A},\cite{Nehra-11-A},\cite{Nhat-13-A},\cite{Kumari-13-A},\cite{Hamdi-10-Aug-A},\cite{Bai-10-Jun-P},\cite{Bok-12-Jun-P},\cite{Mathecken-12-Sep-P}\\
\hline
\multicolumn{2}{ |c| }{Multiple Antenna} &
 \cite{Baum-11-A},\cite{Eldemerdash-13-A} &
 \cite{Gottumukkala-12-A},\cite{Rugini-11-Dec-P},\cite{Shahriar-12-Jun-P} \\
\hline
\multirow{3}{*}{Cooperative}
& AF-OWRN &
 &
\cite{Berger-10-May-P},\cite{Kikuchi-13-Jun-P}  \\
\cline{2-4}
& TWRN &
\cite{Gao-14-A} &
\\
\hline
\multirow{6}{*}{Multiuser / Multicell  }
& SC-FDMA uplink &
\multicolumn{2}{c|}{\cite{Gomaa-14-A}} \\
\cline{2-4}

& OFDMA uplink &
\multicolumn{2}{c|}{\cite{Raghunath-11-A},\cite{Choi-13-A},\cite{Rui-13-A},\cite{Hashemizadeh-11-Sep-P},\cite{Kotzsch-10-Apr-P},\cite{Masucci-13-Sep-P} } \\
\cline{2-4}
& CDMA &
&
\cite{Ahmed-13-A} \\
\cline{2-4}
& Cognitive Radio &
\cite{Pour-10-A} &
\cite{Chen-12-Jun-A},\cite{Ghasabeh-12-A},\cite{Jayaprakasam-10-May-P},\cite{Blad-12-Jun-P},\cite{Yu-10-May-P}\\
\cline{2-4}
& Distributed Multiuser &
&
\cite{Zhu-14-Jun-P}\\
\cline{2-4}
& Multicell Interference &
&
\cite{Sheu-13-Apr-P},\cite{Kotzsch-10-Apr-P},\cite{Pec-13-A} \\
\hline
\multirow{2}{*}{ Other  }
& UWB &
\multicolumn{2}{ c| }{\cite{Sen-11-May-P}} \\
\cline{2-4}
& Spread Spectrum &
\cite{Liu-11-J-A}&
\cite{Chen-11-A}\\
\hline
 \end{tabular}
\label{tab:Coop_MC2}
\end{table*}

\section{SISO Systems}\label{sec:siso}

\subsection{Single-carrier SISO communication systems}

\begin{table*}[t]
\vspace{1cm}
\caption{Summary of synchronization research in single-carrier SISO communication systems} \centering
\begin{tabular}{|l|l|l|l|l|l|l|l|l|} \hline
Article & Channel Model & CSI Req. & CE &  Blind/Pilot & Est/Comp & TO/CFO  & Bound & Comments  \\ \hline
\cite{Lin-14-A} & Freq. sel. & No & No & Blind & Both & Both & Yes & FDE \\
\cite{Shaw-13-A} & AWGN & N/A  & N/A & Pilot & Est & Both & Yes & TS design \\
\cite{Lin-13-A} & Freq. flat & Yes & No & Pilot & Est & Both & Yes &  \\
\cite{Wu-11-A} & AWGN & N/A & N/A & Pilot & Both & CFO & No & Turbo coding \\
\cite{Yin-14-A} & AWGN & N/A & N/A & Blind & Comp & CFO & N/A & hardware implementation \\
\cite{Rabiei-11-A} & AWGN & N/A & N/A & Blind & Comp & CFO & N/A & \\
\cite{Colonnese-10-A} & AWGN &  N/A & N/A  & Blind & Est & CFO & Yes &  \\
\cite{Bai-10-A} & AWGN &  N/A & N/A  & Pilot & Est & CFO & Yes &  \\
\cite{Colonnese-12-A} & AWGN &  N/A & N/A  & Blind & Est & CFO & Yes &  \\
\cite{Oh-13-A} & Freq. sel. & No & No & Pilot & Est & Both & No &  \\
\cite{Man-13-A}  & AWGN & N/A & N/A & Blind & Both & CFO & Yes & LDPC coding \\
\cite{Pedrosa-10-A} & Freq. sel. & Yes & No & Blind & Both & CFO & N/A & FDE \\
\cite{Pan-12-A} & Freq. sel. & No & No & Pilot & Both & CFO & No & IQ imbalance \\
\cite{Sahinoglu-11-A} & AWGN & N/A & N/A & Blind & Est & TO & No & CFO presence \\
\cite{Huh-10-A} & Freq. flat & Yes & No & Pilot & Both & Both & No & PHN \\
\cite{Man-13-M-A} & AWGN &  N/A & N/A  & Blind & Both & TO & Yes & LDPC coding \\
\cite{Kim-10-Aug-A} & Freq. sel. & No & Yes & Pilot & Both & CFO & No & PHN \\
\cite{Hosseini-13-A} & AWGN & N/A & N/A & Pilot & Est & Both & Yes & TS design \\
\cite{Zhang-14-Feb-A} & Freq. sel. & Yes & Yes & Pilot & Both & CFO & No &  FDE, IQ imbalance \\
\cite{Wu-11-Jul-A} & AWGN & N/A & N/A & Blind  & Both & TO & No & Turbo coding  \\
\cite{Gong-12-A} & AWGN & N/A & N/A  & Pilot & Est & CFO & No &  \\
\cite{Hosseini-13-Dec-A} & AWGN &  N/A & N/A   & Pilot & Both & Both & No &  \\
\cite{Nasir-12-PT-A} & AWGN &  N/A & N/A   & Blind & Both & Both & Yes &  \\
\cite{Ramakrishnan-12-A} & AWGN &  N/A & N/A & Blind & Both & CFO & No &  \\
\cite{Dobre-12-A} & AWGN &  N/A & N/A & Blind & Comp & Both & No & Phase offset \\
\cite{Punchihewa-10-A} & Freq. sel. &  No & No & Blind & Comp & Both & No & \\
\cite{Nasir-11-May-P} & Freq. flat & No & Yes & Blind & Both & CFO & No &  \\
\cite{Elgenedy-13-Jun-P} & AWGN &  N/A & N/A & Blind & Est. & CFO & No & Symbol rate est  \\
\cite{Yan-11-Jun-P} & AWGN & N/A & N/A & Blind & Est & CFO & No & Doppler-rate est \\
\cite{Yan-10-May-P} & AWGN & N/A & N/A & Pilot & Comp & CFO & No & SNR est \\
\cite{Obara-10-May-P} & Freq. sel. & Yes & No & Pilot & Comp & TO & No & FDE \\
\cite{Bhatti-11-Sep-P} & AWGN & N/A & N/A  & Pilot & Both & CFO & No & PHN \\
\cite{Zhao-13-Jun-P} & Freq. flat & No & Yes & Blind & Both & CFO & No &  \\
\cite{Popp-13-Sep-P} & AWGN &  N/A & N/A & Blind  & Both & CFO & Yes &  \\
\cite{Bao-12-May-P} &  AWGN &  N/A & N/A & Blind  & Both & TO & No &  \\
\cite{Wang-10-May-P} & AWGN & N/A & N/A  & Pilot & Est & CFO & No &  \\
\cite{Inserra-10-Sep-P} & Freq. flat & Yes& No  & Blind & Comp & Both & No & DoA estimation \\
\cite{Peng-14-Art}   &   AWGN & N/A & N/A & Blind  & Est & CFO & Yes &  \\
\hline
 \end{tabular}
\label{tab:SISO_SC}
\end{table*}

\subsubsection{System Model} In single-carrier single-input-single-output (SISO) systems, a single antenna transmitter communicates with a single antenna receiver and the information is modulated over a single carrier.

The transmitter is assumed to communicate with the receiver through an additive white Gaussian noise (AWGN) or frequency-flat/frequency-selective fading channel. In frequency-flat fading, the coherence bandwidth of the channel is larger than the bandwidth of the signal. Therefore, all frequency components of the signal experience the same magnitude of fading. On the other hand, in frequency-selective fading, the coherence bandwidth of the channel is smaller than the bandwidth of the signal. Therefore, different frequency components of the signal experience uncorrelated fading.

At the receiver end, the effect of channel can be \textit{equalized} either in the time domain or the frequency domain. Time domain equalization is a simple single tap or multi-tap filter. In frequency domain equalization, also referred to as single-carrier frequency domain equalization (SC-FDE), frequency domain equalization is carried out via the fast Fourier transform (FFT) and inverse fast Fourier transform (IFFT) operations at the receiver. Moreover, in SC-FDE, cyclic prefix is appended at the start of the transmission block to take care of the multipath channel effect, such that the length of the cyclic prefix is larger than the multipath channel and the transmission block length is equal to the size of FFT. SC-FDE can be thought of as a single carrier version of orthogonal frequency division multiplexing.

\subsubsection{Synchronization Challenge} The received signal at the receiver is affected by a single timing offset (TO) and a single carrier frequency offset (CFO). The receiver has to estimate these parameters and compensate for their effects from the received signal in order to decode it. The receiver may or may not have the knowledge of channel state information (CSI). In case of no CSI availability, a receiver has to carry out channel estimation (CE) in addition to TO or CFO estimation. The estimation of TO and CFO can be achieved using pilots or by blind methods. For pilot-based estimation, a transmitter sends known training signal (TS) to the receiver before sending the actual data. For blind estimation, a receiver estimates the synchronization parameters using unknown received data. Note that there exists coupling between channel and CFO estimation and their joint estimation is helpful to achieve the best estimation accuracy for these parameters~\cite{Mehr-11-A,Nasir-12-A}.


\subsubsection{Literature Review}
The summary of the research carried out to achieve timing and carrier synchronization in single-carrier SISO communication systems is given in Table \ref{tab:SISO_SC}:
 \begin{enumerate}

 \item The estimation or compensation of timing offset alone and frequency offset alone is studied in \cite{Wu-11-A,Yin-14-A,Rabiei-11-A,Colonnese-10-A,Bai-10-A,Colonnese-12-A,Man-13-A,Pedrosa-10-A,Pan-12-A,Kim-10-Aug-A,Zhang-14-Feb-A,Gong-12-A,Ramakrishnan-12-A,Nasir-11-May-P,Elgenedy-13-Jun-P,Yan-11-Jun-P,Yan-10-May-P,Bhatti-11-Sep-P,Zhao-13-Jun-P,Popp-13-Sep-P,Wang-10-May-P}
and \cite{Sahinoglu-11-A,Man-13-M-A,Wu-11-Jul-A,Obara-10-May-P,Bao-12-May-P,Peng-14-Art}, respectively.

\item Joint timing and carrier synchronization is studied in   \cite{Lin-14-A,Shaw-13-A,Lin-13-A,Oh-13-A,Huh-10-A,Hosseini-13-A,Hosseini-13-Dec-A,Nasir-12-PT-A,Dobre-12-A,Punchihewa-10-A,Inserra-10-Sep-P}.
 \end{enumerate}

The categorized papers differ in terms of channel model, channel estimation requirements or pilot/training requirements. They also differ in whether proposing estimation, compensation, joint channel estimation or estimating lower bound. Further details or differences among these papers are provided in the last column of Table~\ref{tab:SISO_SC}, which further indicates whether any additional parameter such as phase noise (PHN), IQ imbalance, signal-to-noise-ratio (SNR) estimation, or direction of arrival (DoA) estimation, is considered. Moreover, whether training sequence (TS) design or hardware implementation is taken into consideration is also labeled in this table.

\subsubsection{Summary}
Timing and carrier synchronization for single-carrier SISO communication systems is a very well researched topic. The majority of the papers in Table \ref{tab:SISO_SC} are published before 2012. Generally, it is not possible to identify the best pilot-based and best blind-based estimator since the papers have widely different system model assumptions. Future work in this area should compare the performance of their proposed solutions to existing work in Table~\ref{tab:SISO_SC} with similar assumptions to make clear how the state-of-the-art is advancing.

\subsection{Multi-carrier SISO communication systems}

\begin{table*}[p]
\vspace{1cm}
\caption{Summary of research in multi-carrier SISO communication systems considering carrier synchronization.} \centering
\begin{tabular}{|C{5.5cm}|l|l|l|l|l|l|l|} \hline \hline
Article             & Channel Model & CSI Req. & CE &  Blind/Pilot  & Est/Comp & TO/CFO  & Bound   \\ \hline
\cite{Lin-13-May-A},
\cite{Xu-12-A},
\cite{Tseng-13-A},
\cite{Wu-10-A},
\cite{Lu-12-A},
\cite{Wang-10-Mar-A},
\cite{Zhu-13-FQ-A},
\cite{Xu-13-Jun-A},
\cite{Ketseoglou-10-A},
\cite{Lottici-10-A},
\cite{Li-13-Jun-P},
\cite{Zhang-10-May-P},
\cite{Gao-10-Sep-P},
\cite{Guillaud-13-Apr-P},
\cite{Andgart-10-Sep-P},
\cite{Gregorio-10-Apr-P},
\cite{Liu-12-Sep-P}    & Freq. sel. & Yes & Yes & Pilot & Both & CFO & No   \\ \hline
\cite{Cai-11-A},
\cite{Liu-11-A},
\cite{Salim-14-A},
\cite{Rahimi-14-A},
\cite{Morelli-10-A-A},
\cite{Liu-10-May-A},
\cite{Ishaque-13-Sep-P},
\cite{Simon-12-A}      & Freq. sel. & Yes & Yes & Pilot & Both & CFO & Yes  \\  \hline
\cite{Yang-12-A}       & Freq. sel. & Yes & Yes & Pilot & Est  & CFO  & No  \\ \hline
\cite{Nguyen-10-A},
\cite{Cvetkovic-13-A},
\cite{Carvajal-13-A}   & Freq. sel. & Yes & Yes & Pilot & Est  & CFO  & Yes  \\ \hline
\cite{Punchihewa-11-A},
\cite{Dweik-10-A},
\cite{Lmai-14-A},
\cite{Oh-11-A},
\cite{Xu-10-A-A},
\cite{Li-10-May-P}     & Freq. sel. & Yes & No & Blind & Est & CFO & No  \\ \hline
\cite{Kiayani-11-Sep-P}& Freq. sel. & Yes& Yes & Blind & Both & CFO & No  \\   \hline
\cite{Lin-13-Jun-P}    & Freq. sel. & No & No & Blind & Est & CFO & No  \\ \hline
\cite{Lin-10-A}        & Freq. sel. & No & No & Pilot & Est & CFO & No  \\ 
\cite{Zhang-13-A}      & Freq. sel. & Yes & No & Blind & Both & CFO & No  \\ \hline
\cite{Dweik-10-N-A},
\cite{Jeon-11-A}       & Freq. sel. & Yes & No & Blind & Both & CFO & Yes  \\ \hline
\cite{Cai-10-A},
\cite{Shim-10-A},
\cite{You-10-A},
\cite{You-10-O-A},
\cite{Morelli-13-A},
\cite{Bayon-10-A},
\cite{Bai-13-A},
\cite{Su-13-A},
\cite{Wang-12-A},
\cite{Chu-14-A},
\cite{Lee-11-A},
\cite{Marchetti-14-A},
\cite{Kang-10-A},
\cite{Chang-12-Sep-P}   & Freq. sel. & Yes & No & Pilot & Est & CFO & No \\ \hline
\cite{Bai-12-A},
\cite{Bassiouni-13-A},
\cite{Yuan-14-A},
\cite{Lee-13-A},
\cite{Lim-13-A},
\cite{Morelli-12-A},
\cite{Kim-11-Jun-A},
\cite{Wang-13-A},
\cite{Morelli-14-Apr-A},
\cite{Morelli-13-Feb-A},
\cite{Morelli-10-Dec-P},
\cite{Kume-12-Jun-P},
\cite{Dainelli-11-Jun-P}& Freq. sel. & Yes & No & Pilot & Est  & CFO & Yes   \\ \hline
\cite{Wang-14-A}       & Freq. sel. & Yes & No & Pilot & Both & CFO & Yes   \\ \hline
\cite{He-11-A},
\cite{Tsai-10-May-P},
\cite{Guo-11-May-P},
\cite{Miyashita-12-May-P}& Freq. sel. & Yes & No & Pilot & Both & CFO & No   \\ \hline
\cite{Zhang-11-May-P}    & AWGN       & Yes & No & Blind & Both & CFO & No   \\ \hline
\cite{Xie-12-A},
\cite{Dwivedi-12-A},
\cite{Zhang-12-May-A},
\cite{Ma-12-A}         & Freq. sel. & Yes & No & Pilot & Comp & CFO & No  \\ \hline
\cite{Shi-10-A},
\cite{Yeh-12-Dec-P},
\cite{Smida-11-Dec-P},
\cite{Wen-11-Dec-P}    & Freq. sel. & Yes & No & Blind & Comp & CFO & No  \\ \hline
\cite{Hong-11-Sep-P}   & Freq. sel. & Yes & Yes & Pilot & Comp & CFO & No  \\ \hline
\cite{Yang-14-A}       & Freq. sel. & No & Yes & Pilot & Est  & CFO & No  \\ \hline
\cite{Lanlan-10-A}     & Freq. sel. & Yes & Yes& semiblind & Both  & CFO & Yes  \\ \hline
\hline
 \end{tabular}
\label{tab:SISO_MC_CFO}
\vspace{1cm}
\caption{Summary of research in multi-carrier SISO communication systems considering timing synchronization.} \centering
\begin{tabular}{|l|l|l|l|l|l|l|l|l|} \hline
Article             & Channel Model & CSI Req. & CE  &  Blind/Pilot  & Est/Comp & TO/CFO  & Bound & Comments \\ \hline
\cite{Tanhaei-11-A} &   Freq. sel. & Yes      & Yes &     Pilot     & Both     &     TO  &  No   & DVB-T system \\
\cite{Sheng-10-A}   &    Freq. sel. & Yes      & Yes &     Pilot     & Est    &     TO  &  No   & subspace based est\\
\cite{Dweik-12-A}   &    Freq. sel. & No      & No  &    Blind      & Both    &     TO  &  No   & subspace based est\\
\cite{Zhang-12-Feb-A} &  Freq. sel. & No      & No  &    Pilot     & Both    &     TO  &  Yes   & autocorrelation based estimation\\
\cite{Mohebbi-14-A} &    Freq. sel. & No      & No  &    Pilot     & Both    &     TO  &  Yes    & fourth order statistics\\
\cite{Kung-12-A}    &    Freq. sel. & Yes      & Yes &     Pilot     & Both     &     TO  &  No   & ML estimation\\
\cite{Yang-12-Dec-P}&    Freq. sel. & No      & No &     Pilot      & Est     &     TO  &  Yes   & SNR estimation\\
\cite{Wang-11-Sep-P}&    Freq. sel. & No      & Yes &     Blind      & Both    &     TO  &  No  & throughput computation\\
\cite{Chang-10-Sep-P}&    Freq. sel. & No      & No &     Pilot     & Both    &     TO  &  No  & \\
\cite{Pan-11-May-P}  &  Freq. sel.  & No    & No &     Pilot     & Est    &     TO  &  No & \\
\cite{Chan-13-Sep-P} & Freq. sel.  & No    & No &     Pilot     & Est    &     TO  &  No & \\
\cite{Liu-14-Nov-Art}& Freq. sel. & No    & No     & Pilot      & Est    & TO  & No  & immune to CFO \\
\hline
 \end{tabular}
\label{tab:SISO_MC_TO}
\vspace{1cm}
\caption{Summary of research in multi-carrier SISO communication systems considering joint timing and carrier synchronization.} \centering
\begin{tabular}{|l|l|l|l|l|l|l|l|} \hline
Article             & Channel Model & CSI Req. & CE &  Blind/Pilot  & Est/Comp & TO/CFO  & Bound   \\ \hline
\cite{Li-12-A},
\cite{Li-12-Dec-A},
\cite{Thein-13-Sep-P},
\cite{Thein-13-Jun-P},
\cite{Welden-12-A}  & Freq. sel. & Yes & Yes & Pilot & Both & Both & No   \\
\cite{Wang-11-Jun-P}& Freq. sel. & Yes & Yes & Pilot & Est & Both & No   \\  
\cite{Sanguinetti-10-A}   & Freq. sel. & Yes & No  & Pilot & Both & Both & Yes \\ 
\cite{Long-11-A},
\cite{Rotoloni-12-A},
\cite{Choi-10-A}    & Freq. sel. & Yes & No  & Blind & Est  & Both & Yes  \\
\cite{Liu-10-A},
\cite{Wiegand-12-Sep-P}  & Freq. sel. & Yes & No  & Pilot & Both & Both & No   \\
\cite{Nasraoui-12-Apr-P}  & AWGN & N/A & N/A  & Pilot & Both & Both & No   \\ 
\cite{Ziabari-11-A},
\cite{Lifeng-10-A},
\cite{Viemann-10-A},
\cite{Xu-10-Dec-P}  & Freq. sel. & Yes & No  & Pilot & Est & Both  & No   \\
\cite{Chin-14-May-A},
\cite{Xu-13-Dec-A},
\cite{Udupa-13-Jun-P}& Freq. sel. & Yes & No  & Pilot & Est & Both  & Yes   \\
\cite{Chin-14-A}    & Freq. sel. & Yes & No  & Blind & Both & Both & No \\ 
\cite{Punchihewa-10-A} & Freq. sel. &  No & No & Blind & Comp & Both & No  \\
\hline
 \end{tabular}
\label{tab:SISO_MC_TO_CFO}
\end{table*}

\subsubsection{System Model} In multi-carrier systems, information is modulated over multiple carriers. The well known multi-carrier system is based upon orthogonal frequency division multiplexing (OFDM).\footnote{The system model and the synchronization challenge for other types of multi-carrier systems, e.g., filter bank multi-carrier (FBMC) systems etc. are not considered in this paper. Their details can be found in the papers identified in Section \ref{sec:SISO-MC-LR}.} In OFDM systems, at the transmitter side, an IFFT is applied to create an OFDM symbol and a cyclic prefix is appended to the start of an OFDM symbol. At the receiver, the cyclic prefix is removed and an FFT is applied to the received OFDM symbol. Note that frequency domain processing greatly simplifies receiver processing. The length of the cyclic prefix is designed to be larger than the span of the multipath channel. The portion of the cyclic prefix which is corrupted due to the multipath channel from the previous OFDM symbols is known as the \textit{inter symbol interference (ISI) region}. The remaining part of the cyclic prefix which is not affected by the multipath channel is known as the ISI-free region. Note that cyclic prefix can mainly remove the ISI and proper design of the cyclic prefix length has been a design issue under research.

\subsubsection{Synchronization Challenge}\label{sec:SISO-MC-SC} In OFDM systems, the presence of TO affects the system performance in a different way as compared to single-carrier systems:
\begin{enumerate}
\item
If the TO lies within the ISI-free region of the cyclic prefix, the orthogonality among the subcarriers is not destroyed and the timing offset only introduces a phase rotation in every subcarrier symbol. For a coherent system, this phase rotation is compensated for by the channel equalization scheme, which views it as a channel-induced phase shift. 
\item
If the TO is outside the limited ISI-free region, the orthogonality amongst the subcarriers is destroyed by the resulting ISI and additional inter carrier interference (ICI) is introduced. 
\end{enumerate}

\noindent Thus, the objective of timing synchronization in OFDM systems, unlike in single-carrier systems, is to identify the start of an OFDM symbol within the ISI-free region of the cyclic prefix.

The presence of CFO in OFDM systems attenuates the desired signal and introduces ICI since the modulated carrier is demodulated at an offset frequency at the receiver side. In OFDM systems, CFO is usually represented in terms of subcarrier spacings and can be divided into an integer part (integer number less than the total number of subchannels) and a fractional part (within $\pm \frac{1}{2}$ of subcarrier spacing). If the CFO is greater than the subcarrier spacing, a receiver has to estimate and compensate for both integer and fractional parts of the normalized CFO. 

The synchronization in OFDM systems can be performed either in the time domain or the frequency domain depending upon whether the signal processing is executed pre-FFT or post-FFT at the receiver, respectively.

\subsubsection{Literature Review} \label{sec:SISO-MC-LR}
The summary of the research carried out to achieve carrier synchronization, timing synchronization, and joint timing and carrier synchronization in multi-carrier SISO communication systems is given in Tables \ref{tab:SISO_MC_CFO}, \ref{tab:SISO_MC_TO}, and \ref{tab:SISO_MC_TO_CFO}, respectively. Their details are given below.
\begin{enumerate}[(a)]
\item Carrier Synchronization: \\ The papers studying carrier synchronization in multi-carrier SISO communication systems are listed in Table~\ref{tab:SISO_MC_CFO}. It can be observed that there are groups of papers which consider the same channel model and the same requirement for CSI and training. Also, they consider the same problem in terms of estimation or compensation. In the following, we describe how these papers differ within their respective groups.
\begin{enumerate}[(i)]
\item \emph{Pilot based CFO estimation and compensation with channel estimation:} \\
The papers here can be grouped into two categories. The first group does not provide an estimation error lower bound \cite{Lin-13-May-A,Xu-12-A,Tseng-13-A,Wu-10-A,Lu-12-A,Wang-10-Mar-A,Zhu-13-FQ-A,Xu-13-Jun-A,Ketseoglou-10-A,Lottici-10-A,Li-13-Jun-P,Zhang-10-May-P,Gao-10-Sep-P,Guillaud-13-Apr-P,Andgart-10-Sep-P,Gregorio-10-Apr-P,Liu-12-Sep-P}. In addition to carrier synchronization, \cite{Lin-13-May-A} proposes to achieve seamless service in vehicular communication and also considers road side unit selection, \cite{Xu-12-A} considers CFO tracking assuming constant modulus based signaling, \cite{Tseng-13-A} considers concatenated precoded OFDM system, \cite{Wu-10-A} proposes MMSE based estimation, \cite{Lu-12-A} proposes hard decision directed based CFO tracking, \cite{Wang-10-Mar-A} considers phase rotated conjugate transmission and receiver feedback, \cite{Zhu-13-FQ-A} considers hardware implementation with IQ imbalance and power amplifier nonlinearity, \cite{Xu-13-Jun-A} considers hexagonal multi-carrier transmission system and a doubly dispersive channel, \cite{Ketseoglou-10-A} considers maximum a posteriori expectation-maximization (MAP-EM) based Turbo receiver, \cite{Lottici-10-A} considers an FBMC system, \cite{Li-13-Jun-P} considers aerial vehicular communication, \cite{Zhang-10-May-P} proposes noise variance estimation and considers EM algorithm, \cite{Gao-10-Sep-P} considers SFO estimation, \cite{Guillaud-13-Apr-P} considers hardware implementation, \cite{Andgart-10-Sep-P} proposes estimation of the CFO over a wide range of offset values, \cite{Gregorio-10-Apr-P} considers IQ imbalance and phase noise distortion, and \cite{Liu-12-Sep-P} considers Doppler spread in a mobile OFDM system.

The second group of papers provides an estimation error lower bound on obtaining the CFO
\cite{Cai-11-A,Liu-11-A,Salim-14-A,Rahimi-14-A,Morelli-10-A-A,Liu-10-May-A,Ishaque-13-Sep-P,Simon-12-A}. In addition to carrier synchronization, \cite{Cai-11-A} considers IQ imbalance, \cite{Salim-14-A} proposes an extended Kalman filter (EKF) based estimator in the presence of phase noise, \cite{Rahimi-14-A} proposes an ML estimator and considers an FBMC system, \cite{Morelli-10-A-A} considers SFO estimation and synchronization, \cite{Liu-10-May-A} considers ML based frequency tracking, and \cite{Ishaque-13-Sep-P} considers IQ imbalance and its estimation.
\item \emph{Pilot based CFO estimation with channel estimation:} \\
The papers \cite{Nguyen-10-A,Cvetkovic-13-A,Carvajal-13-A} fall under this category. In addition to carrier synchronization, \cite{Cvetkovic-13-A} proposes computationally efficient, single training sequence based least squares estimation, \cite{Nguyen-10-A} considers doubly-selective channel estimation, and \cite{Carvajal-13-A} proposes an EM-based ML estimator and also considers the presence of phase noise.
\item \emph{Blind CFO estimation with no channel estimation:} \\
The papers here can be grouped into two categories. The first group does not provide an estimation lower bound \cite{Punchihewa-11-A,Dweik-10-A,Lmai-14-A,Oh-11-A,Xu-10-A-A,Li-10-May-P}. In addition to carrier synchronization, \cite{Punchihewa-11-A} considers a cognitive radio network and the algorithm applies even if timing offset is unknown, \cite{Dweik-10-A} considers time-varying channels and Doppler frequency, \cite{Lmai-14-A} and \cite{Oh-11-A} consider constant modulus based signaling, \cite{Xu-10-A-A} considers cyclic correlation based estimation, the estimator proposed by \cite{Li-10-May-P} is based on minimum reconstruction error, and \cite{Simon-12-A} proposes an EM based estimator considering very high mobility.

The second group of papers provides an error lower bound on CFO estimation \cite{Dweik-10-N-A,Jeon-11-A}. In addition to carrier synchronization, \cite{Dweik-10-N-A} proposes a Viterbi-based estimator and \cite{Jeon-11-A} proposes CFO estimation using single OFDM symbol and provides closed-form expression for the CFO estimate using property of the cosine function.
\item \emph{Pilot based CFO estimation with no channel estimation:} \\
The papers here can be grouped into two categories. The first group of papers does not provide an estimation error lower bound \cite{Cai-10-A,Shim-10-A,You-10-A,You-10-O-A,Morelli-13-A,Bayon-10-A,Bai-13-A,Su-13-A,Wang-12-A,Chu-14-A,Lee-11-A,Marchetti-14-A,Kang-10-A,Chang-12-Sep-P}. In addition to carrier synchronization, the CFO estimation algorithm proposed by \cite{Cai-10-A} is valid even if timing offset and channel length is unknown, the algorithm proposed by \cite{Shim-10-A} estimates integer frequency offset, the algorithm proposed by \cite{You-10-A} estimates sampling frequency offset in addition to CFO, \cite{You-10-O-A} estimates IFO for OFDM-based digital radio mondiale plus system, \cite{Morelli-13-A} considers IQ imbalance and direct-conversion receivers, \cite{Bayon-10-A} considers CFO tracking in digital video broadcasting (DVB-T) system, \cite{Bai-13-A} proposes ML based estimation, \cite{Su-13-A} and \cite{Chu-14-A} propose IFO estimation with cell sector identity detection in long term evolution systems, \cite{Wang-12-A} considers IQ imbalance and hardware implementation, \cite{Lee-11-A} also considers SFO estimation, \cite{Marchetti-14-A} proposes ML based estimation and considers the design of pilot pattern, the estimation algorithm in \cite{Kang-10-A} is robust to the presence of Doppler shift and \cite{Chang-12-Sep-P} considers IQ imbalance and its estimation.

The second group of papers provides an error lower bound on CFO estimation \cite{Bai-12-A,Bassiouni-13-A,Yuan-14-A,Lee-13-A,Lim-13-A,Morelli-12-A,Kim-11-Jun-A,Wang-13-A,Morelli-14-Apr-A,Morelli-13-Feb-A,Morelli-10-Dec-P,Kume-12-Jun-P,Dainelli-11-Jun-P}. In addition to carrier synchronization, \cite{Bai-12-A} derives CRLB for the general case where any kind of subcarriers, e.g., pilot, virtual, or data subcarriers may exist, \cite{Bassiouni-13-A} considers eigen-value based estimation, \cite{Yuan-14-A} considers subspace based channel estimation with hardware implementation and SNR detection, \cite{Lee-13-A} considers IFO estimation and training sequence design, \cite{Lim-13-A} considers Gaussian particle filtering based estimation, \cite{Morelli-12-A} proposes both IFO and FFO estimation while also considering IQ imbalance and a direct conversion receiver structure, \cite{Kim-11-Jun-A} considers SFO estimation while proposing ML based estimation, \cite{Wang-13-A} proposes multiple signal classification or a subspace based estimation method, \cite{Morelli-14-Apr-A} proposes an estimator based on the space-alternating generalized expectation-maximization (SAGE) algorithm and considers IQ imbalance, \cite{Morelli-13-Feb-A} proposes SNR and noise power estimation, \cite{Morelli-10-Dec-P} and \cite{Kume-12-Jun-P} consider IQ imbalance, and \cite{Dainelli-11-Jun-P} considers doubly selective fading channels.
\item \emph{Pilot based CFO estimation and compensation with no channel estimation:} \\
The papers \cite{He-11-A,Tsai-10-May-P,Guo-11-May-P,Miyashita-12-May-P} fall under this category. In addition to carrier synchronization, \cite{He-11-A} proposes training sequence design in DVB-T2 system, \cite{Tsai-10-May-P} considers frequency domain pilot signaling, \cite{Guo-11-May-P} also considers SFO estimation, and \cite{Miyashita-12-May-P} considers IQ imbalance and its estimation.
\item \emph{Pilot based CFO compensation with no channel estimation:} \\
The papers \cite{Xie-12-A,Dwivedi-12-A,Zhang-12-May-A,Ma-12-A} fall under this category. The differences among them are that in addition to carrier synchronization, \cite{Dwivedi-12-A} proposes repeated correlative coding for mitigation of ICI, \cite{Xie-12-A} proposes training sequence design, \cite{Zhang-12-May-A} considers cell identification in long term evolution (LTE) system, and \cite{Ma-12-A} considers detection of primary synchronization signal in LTE systems.
\item \emph{Blind CFO compensation with no channel estimation:} \\
The papers \cite{Shi-10-A,Yeh-12-Dec-P,Smida-11-Dec-P,Wen-11-Dec-P} fall under this category. The differences among them are that in addition to carrier synchronization, \cite{Shi-10-A} and \cite{Yeh-12-Dec-P} propose EKF based algorithm and space time parallel cancellation schemes, respectively, to cancel out inter carrier interference due to CFO, \cite{Smida-11-Dec-P} proposes reduction of peak interference to carrier ratio, and \cite{Wen-11-Dec-P} considers IQ imbalance.
\end{enumerate}

\item Timing Synchronization: \\Compared to the categorized papers for carrier synchronization in Table~\ref{tab:SISO_MC_CFO}, the categorized papers for timing synchronization in Table \ref{tab:SISO_MC_TO} have greater similarity. The major differences are found in terms of channel estimation requirement, pilot/training requirement, and lower bounds on the estimation performance. Further details are provided in the last column of Table \ref{tab:SISO_SC}, which also indicates if any additional parameter, e.g., SNR estimation is considered.

\item Joint Timing and Carrier Synchronization: \\The papers studying joint timing and carrier synchronization in multi-carrier SISO communication systems are listed in Table \ref{tab:SISO_MC_TO_CFO}. It can be observed that there are groups of papers which consider the same channel model and the same requirement for CSI and training. Also, they further consider the same problem in terms of estimation or compensation. In the following, we describe how these papers differ within their respective groups.
\begin{enumerate}[(i)]
\item \emph{Pilot based TO and CFO estimation and compensation with channel estimation:} \\
The papers \cite{Li-12-A,Li-12-Dec-A,Thein-13-Sep-P,Thein-13-Jun-P,Welden-12-A} fall under this category. In addition to joint timing and carrier synchronization, \cite{Li-12-A} considers IFO estimation while considering residual timing offset, \cite{Li-12-Dec-A} considers hardware implementation, \cite{Thein-13-Sep-P} considers FBMC system, \cite{Thein-13-Jun-P} considers offset-QAM modulation, and \cite{Welden-12-A} considers decision directed based estimation.
\item \emph{Blind TO and CFO estimation with no channel estimation:} \\
The papers \cite{Rotoloni-12-A,Choi-10-A} fall under this category. In addition to joint timing and carrier synchronization, \cite{Rotoloni-12-A} considers digital video broadcasting (DVB-T2) standard and \cite{Choi-10-A} proposes ML estimation with a time-domain preamble.
\item \emph{Pilot based TO and CFO estimation and compensation with no channel estimation:} \\
The papers \cite{Liu-10-A,Wiegand-12-Sep-P} fall under this category. In addition to joint timing and carrier synchronization, \cite{Liu-10-A} considers time domain synchronous (TDS)-OFDM system which replaces cyclic prefix with a pseudo noise (PN) and thus, proposes PN-correlation based synchronization and \cite{Wiegand-12-Sep-P} considers hardware implementation.
\item \emph{Pilot based TO and CFO estimation with no channel estimation:} \\
The first group of papers does not provide an estimation error lower bound \cite{Ziabari-11-A,Lifeng-10-A,Viemann-10-A,Xu-10-Dec-P}. The differences among them are that in addition to joint timing and carrier synchronization, CFO estimation in \cite{Ziabari-11-A} applies to a wide CFO range, i.e., $\pm 1/2$ the total number of subcarriers width, \cite{Lifeng-10-A} considers phase noise (PN)-sequence based preamble, \cite{Viemann-10-A} considers digital video broadcasting (DVB-T2) system, and \cite{Xu-10-Dec-P} considers blind cyclic prefix length in their algorithm.

The second group of papers provides an estimation error lower bound \cite{Chin-14-May-A,Xu-13-Dec-A,Udupa-13-Jun-P}. In addition to joint timing and carrier synchronization, \cite{Chin-14-May-A} considers doubly selective channel, \cite{Xu-13-Dec-A} considers hexagonal multi-carrier transmission system, and \cite{Udupa-13-Jun-P} proposes training sequence design.
\end{enumerate}
\end{enumerate}

\subsubsection{Summary}
Timing and carrier synchronization for multi-carrier SISO communication systems is still an ongoing topic of research, as evidenced by the large number of published papers. In particular, there is a major emphasis on accurate CFO estimation for different types of systems and often in conjunction with RF impairments such as phase noise and IQ imbalance.

\section{Multi-Antenna Systems}\label{sec:mimo}

\subsection{Single-carrier multi-antenna communication systems}

\subsubsection{System Model} \label{sec:MIMO-SC-SM} In a multi-antenna wireless communication system, data is transmitted across different channels that are modeled either as quasi-static or time varying. The received signal at an antenna is given by a linear combination of the data symbols transmitted from different transmit antennas. In order to achieve multiplexing or capacity gain, independent data is transmitted from different transmit antennas.

A space-time multiple-input-multiple-output (MIMO) decoder can be used to decode the signal from multiple antenna streams. On the other hand, in order to achieve diversity gain, the same symbol weighted by a complex scale factor may be sent over each transmit antenna. This latter scheme is also referred to as MIMO beamforming \cite{Goldsmith-B-05}. Depending on the spatial distance between the transmit or receive antennas, which may differ for line-of-sight (LOS) and non-LOS propagation, the antennas may be equipped with either their own oscillators or use the same oscillator. 

Depending on the number of antennas at the transmitter and the receiver, multi-antenna systems can be further categorized into MIMO systems, multiple-input-single-output (MISO) systems, or single-input-multiple-output (SIMO) systems. Further, if the antennas at the transmitter side are not co-located at a single device, such a system is referred to as a distributed-MIMO system, i.e., multiple distributed transmitters simultaneously communicate with a single multi-antenna receiver.

\begin{table*}[t]
\vspace{1cm}
\caption{Summary of synchronization research in single-carrier multi-antenna communication systems} \centering
\begin{tabular}{|l|l|l|l|l|l|l|l|l|l|l|} \hline
Article & System & Fading & CSI Req. & CE &  Blind/Pilot & Est/Comp & TO/CFO  & Bound & Oscillators (Tx/Rx) & Comments  \\ \hline
\cite{Jiang-13-A-A} & Virtual MIMO & Freq. flat & Yes & No & Pilot & Both & CFO & No & single/multiple & Codebook design \\
\cite{Zhang-10-N-A} & SC-FDMA MIMO & Freq. sel. & Yes & No & Pilot & Both & CFO & No & single/single & SFBC, PHN \\
\cite{Zhang-10-May-A} & MIMO & Freq. sel. & No & Yes & Pilot & Both & CFO & No & multiple/multiple & FDE \\
\cite{Du-13-A} & MIMO & Freq. flat & No & Yes & Pilot & Both & CFO & Yes & multiple/multiple & \\
\cite{Marey-12-A} & MIMO & Freq. flat & No & No & Blind & Comp & Both & N/A & single/single & STBC \\
\cite{Eldemerdash-13-Dec-P} & MISO & Freq. flat & No & No & Blind & Comp & Both & N/A & single/single & STBC \\
\cite{Gao-10-May-P} & SC-FDE MIMO & Freq. sel & No & Yes & Pilot & Both & CFO & No & single/single & IQ imbalance \\
\cite{Sinha-13-Sep-P} & SIMO & Freq. flat & No & No & N/A & Comp & CFO & N/A & single/single & \\
\cite{Yao-11-Mar-P} & MISO WSN & Freq. sel & No & Yes & Pilot & Est & CFO & No & single/single & AOD est. \\
\cite{Nasir-11-A}  & distributed MIMO & Freq. flat & No & Yes & Blind & Both & Both & Yes & multile/single & \\
\cite{Mohammadkarimi-14-Oct-Art} & MISO & Freq. flat & No & No & Blind & Comp & CFO & N/A & multile/single & STBC \\
\hline
 \end{tabular}
\label{tab:MIMO_SC}

\vspace{1cm}
\caption{Summary of synchronization research in multi-carrier multi-antenna communication systems} \centering
\begin{tabular}{|l|l|l|l|l|l|l|l|l|l|l|} \hline
Article & System & Fading   & CSI Req. & CE &  Blind/Pilot & Est/Comp & TO/CFO  & Bound & Tx/Rx Oscillator & Comments  \\ \hline
\cite{Salari-11-A} & MIMO & Freq. sel. & No & Yes & Pilot & Both & Both & Yes & single/single & \\
\cite{Choi-14-A} & MISO & Freq. sel. & Yes & No & N/A & Comp & CFO & N/A & multiple/single & Alamouti coding \\
\cite{Kwon-12-A} & MISO & Freq. sel. & Yes & No & Pilot & Est & CFO & No & single/single & IFO est. \\
\cite{Shim-10-D-A} & MISO & Freq. sel. & Yes & No & Pilot & Est & CFO & No & single/single & IFO est. \\
\cite{Younis-10-A} & MIMO & Freq. sel. & No & No & Blind & Est & CFO & No & single/single & \\
\cite{Liu-10-A-A} & MIMO & Freq. sel. & No & No & Pilot & Est & Both & No & multiple/multiple & \\
\cite{Yu-12-A} & MIMO & Freq. sel. & No & Yes & Pilot & Est & CFO & No & multiple/multiple & \\
\cite{Bannour-12-A} & MIMO & Freq. sel. & No & Yes & Pilot & Est & CFO & Yes & multiple/multiple & Algebraic STC \\
\cite{Amo-13-A} & MIMO & Freq. sel. & No & Yes & Pilot & Both & CFO & Yes & single/single & insufficient CP \\
\cite{Zhang-13-M-A} & MIMO & Freq. sel. & No & Yes & Blind & Both & CFO & Yes & single/single &  \\
\cite{Simon-11-A} & MIMO & Freq. sel. & No & Yes & Pilot & Both & CFO & No & multiple/multiple & time varying channel \\
\cite{Liang-10-A} & MIMO & Freq. sel. & No & Yes & Pilot & Est & CFO & Yes & single/single &  \\
\cite{Zhang-A-A-14} & MIMO & Freq. sel. & No & Yes & Blind & Est & CFO & Yes & single/single & \\
\cite{Nguyen-11-A} & Coded MIMO & Freq. sel. & No & Yes & Pilot & Both & CFO & Yes & single/single & doubly sel. channel\\
\cite{Kim-11-A} & Coded MIMO & Freq. sel. & No & Yes & Semiblind & Both & CFO & No & single/single & \\
\cite{Baek-10-A} & MIMO & Freq. sel. & No & Yes & Pilot & Est & TO & No & single/single & \\
\cite{Jeon-14-A} & MIMO & Freq. sel. & No & Yes & Pilot & Both & CFO & Yes & multiple/single & \\
\cite{Hsu-12-A} & MIMO & Freq. sel. & No & No & Pilot & Both & Both & No & single/single & hardware implementation \\
\cite{Jose-13-A} & MIMO & Freq. sel. & No & Yes & Pilot & Both & Both & Yes & single/single &  \\
\cite{Chung-10-A} & MIMO & Freq. sel. & No & Yes & Pilot & Est & CFO & Yes & single/single & IQ imbalance \\
\cite{Baek-11-A} & MIMO & Freq. sel. & No & Yes & Pilot & Both & TO & No & single/single & \\
\cite{Choi-11-A} & distributed MISO & Freq. flat & Yes & No & N/A & Comp & CFO & N/A & multiple/single & Alamouti coding \\
\cite{Narasimhan-10-A} & MIMO & Freq. sel. & No & Yes & Pilot & Comp & CFO & No & single/single & SFBC, IQ imbalance \\
\cite{Weikert-13-Jun-P} & MIMO & Freq. sel. & No & Yes & Pilot & Est & CFO & No & single/single & IQ imbalance, PHN, SFO \\
\cite{Jiang-13-Dec-P} & MIMO & Freq. sel. & Yes & No & Semiblind & Both & CFO & No & single/single & \\
\cite{Mahesh-12-Apr-P} & MIMO & Freq. flat & No & Yes & Pilot & Est & TO & Yes & single/single  & \\
\cite{Xu-13-Apr-P} & MIMO & Freq. sel. & No & Yes & Pilot & Comp & CFO & No & single/single & \\
\cite{Lei-11-Sep-P} & coded MIMO & Freq. sel. & No & Yes & Pilot & Est & CFO & Yes & single/single & STBC \\
\cite{Wang-11-Sep-VTC-P} & distributed MIMO & Freq. sel. & No & No & Pilot & Est & Both & No & multiple/single & TS design \\
\cite{Luo-11-Sep-P} & MIMO & Freq. sel. & No & Yes & Pilot & Both & CFO & Yes & single/single & TS design, IQ imbalance \\
\cite{Wang-10-Sep-P} & MIMO & Freq. sel. & Yes & No & Pilot & Both & CFO & No & single/single & \\
\cite{Jing-14-Sup-Art} & MIMO & Freq. sel. & Yes & Yes & Pilot & Est & CFO & No & multiple/multiple & time varying channel \\
\hline
 \end{tabular}
\label{tab:MIMO_MC}
\end{table*}

\subsubsection{Synchronization Challenge} In multi-antenna systems, multiple signal streams arrive at the receive antenna from different transmit antennas resulting in \textit{multiple timing offsets (MTOs)}. In some special cases multiple timing offsets actually reduce to a single timing offset, e.g., if multiple antennas are co-located at a single transmitter device, then the transmit filters can be synchronized easily and the multiple signal streams arriving at the receive antenna experience approximately the same propagation delay.

If the transmit antennas are fed through independent oscillators, the received signal at the receive antenna is affected by \textit{multiple carrier frequency offsets (MCFOs)} because of the existence of independent frequency offset between each transmit antenna oscillator and the receive antenna oscillator. On the other hand, if the transmit antennas are equipped with a single oscillator, the received signal at the receive antenna is affected by a single frequency offset. Thus, each receive antenna has to estimate and compensate for a single or multiple timing and frequency offsets, depending on the system model assumptions including Doppler fading.

In the case of distributed antenna systems, the receiver has to estimate and compensate for multiple CFOs and multiple TOs because each distributed transmit antenna is equipped with its own oscillator and multiple signal streams arriving at the receive antenna experience different propagation delays. Thus, in practice, the number of distributed antennas may need to be limited to avoid synchronization and pilot overhead associated with obtaining multiple CFOs and TOs. 

\subsubsection{Literature Review}
The summary of the research carried out to achieve timing and carrier synchronization in single-carrier multi-antenna communication systems is given in Table \ref{tab:MIMO_SC}:
\begin{enumerate}

\item The estimation or compensation of CFO alone is studied in \cite{Jiang-13-A-A,Zhang-10-N-A,Zhang-10-May-A,Du-13-A,Gao-10-May-P,Sinha-13-Sep-P,Yao-11-Mar-P,Mohammadkarimi-14-Oct-Art}.

\item The joint timing and carrier synchronization is studied in~\cite{Marey-12-A,Eldemerdash-13-Dec-P,Nasir-11-A}.
\end{enumerate}

The categorized papers differ in terms of channel model, channel estimation requirement or pilot/training requirement. They also differ in terms of proposing estimation, compensation, joint channel estimation or estimation lower bound. Further details or differences are provided in the last column of Table \ref{tab:SISO_SC}, which indicates if any additional parameter, e.g., phase noise (PHN), IQ imbalance or direction of arrival (DoA) estimation, is considered or if space-time block coding (STBC), space frequency block coding (SFBC), or codebook design is considered.

\subsubsection{Summary} Synchronization in single-carrier multi-antenna communication systems has not received as much attention compared to synchronization in multi-carrier multi-antenna communication systems. This may not be surprising since the latter is adopted in current wireless cellular standards.

\subsection{Multi-carrier multi-antenna communication systems}

\subsubsection{System Model} In multi-carrier multi-antenna systems, the information at each antenna is modulated over multiple carriers. Thus, apart from the IFFT/CP addition and CP removal/FFT operations at each transmit and receive antennas, respectively, the system model for multi-carrier multi-antenna communication systems is similar to the one described for single-carrier multi-antenna systems presented in Section \ref{sec:MIMO-SC-SM}.

\subsubsection{Synchronization Challenge} Similar to single-carrier multi-antenna systems, the signal arriving at the receive antenna can potentially be affected by multiple TOs and multiple CFOs, when the transmit antennas are fed by different oscillators and are distant from one another. Due to multiple carriers, the presence of multiple TOs and multiple CFOs results in strong ISI and ICI. The synchronization challenge is to jointly estimate and compensate for the effect of multiple TOs and multiple CFOs in order to mitigate ISI and ICI and decode the signal from multiple antenna streams.

\subsubsection{Literature Review}
The summary of the research carried out to achieve timing and carrier synchronization in single-carrier multi-antenna communication systems is given in Table \ref{tab:MIMO_MC}:
\begin{enumerate}

\item  The estimation or compensation of TO and CFO alone is studied in \cite{Baek-10-A,Baek-11-A,Mahesh-12-Apr-P} and \cite{Choi-14-A,Kwon-12-A,Shim-10-D-A,Younis-10-A,Yu-12-A,Bannour-12-A,Amo-13-A,Zhang-13-M-A,Simon-11-A,Liang-10-A,Zhang-A-A-14,Nguyen-11-A,Kim-11-A,Jeon-14-A,Chung-10-A,Choi-11-A,Narasimhan-10-A,Weikert-13-Jun-P,Jiang-13-Dec-P,
Xu-13-Apr-P,Lei-11-Sep-P,Luo-11-Sep-P,Wang-10-Sep-P,Jing-14-Sup-Art}, respectively.

\item The joint timing and carrier synchronization is studied in \cite{Salari-11-A,Liu-10-A-A,Hsu-12-A,Jose-13-A,Wang-11-Sep-VTC-P}.

\end{enumerate}

The number of oscillators considered by different papers at the transmitter and receiver, respectively, are given under the ``Tx/Rx Oscillator" column. The categorized papers differ in terms of channel model, channel estimation requirement or pilot/training requirement. They also differ in proposing estimation, compensation, joint channel estimation or estimation lower bound. Further details or differences are provided in the last column of Table \ref{tab:MIMO_MC}, which indicates if additional parameters, e.g., phase noise or IQ imbalance is considered or if STBC, SFBC, coding, or hardware implementation is considered.

\subsubsection{Summary} Compared to the estimation of single TO and single CFO, estimation of multiple timing offsets (MTOs) and multiple carrier frequency offsets (MCFOs) is more challenging, due to pilot design issues, overhead, pilot contamination problem, complexity, and non-convex nature of optimization problems. Joint estimation of MTOs and MCFOs is an important unsolved issue, which has been considered by only a handful of the papers.

\section{Cooperative Communication Systems}\label{sec:cooperative}

\begin{table*}[p]
\vspace{1cm}
\caption{Summary of synchronization research in single carrier cooperative communication systems} \centering
\begin{tabular}{|l|l|l|l|l|l|l|l|l|l|l|l|} \hline
Article & Network & DF/AF & DL & Fading  & CSI Req. & CE &  Blind/Pilot & Est/Comp & TO/CFO  & Bound & Comments  \\ \hline
\cite{Avram-12-Sep-P} & OWRN & QF & Yes & Freq. flat & No & Yes & Pilot & Est & TO & Yes & \\
\cite{Mehr-11-A} & OWRN & Both & No & Freq. flat & No & Yes & Pilot & Both & CFO & Yes &  \\
\cite{Nasir-13-A} & OWRN & AF & No & Freq. flat & No & Yes & Pilot  & Both & Both & Yes & DSTBC \\
\cite{Nasir-12-A} & OWRN & Both & No & Freq. flat & No & Yes & Pilot & Both & Both & Yes & \\
\cite{Li-10-A},\cite{Mehrpouyan-11-C-A} & OWRN & AF & No & Freq. flat & No & Yes & Pilot & Both & TO & Yes & \\
\cite{Yadav-10-May-P} & OWRN & AF & Yes & Freq. sel. & No & No & N/A & Comp & Both & No & DSTBC \\
\cite{Mehrpouyan-10-Dec-P} & OWRN & Both & No & Freq. flat & No & Yes & Pilot & Est & TO & Yes & \\
\cite{Liu-12-A} & OWRN & DF & No & Freq. sel. & No & Yes & Pilot & Est & CFO & Yes & \\
\cite{Nasir-13-AT} & OWRN & DF & No & Freq. flat & No & Yes & Pilot & Both & TO & Yes & TS design \\
\cite{Wang-11-Feb-A} & OWRN & DF & No & Freq. flat & Yes & No & N/A & Comp & CFO & No & DSFBC \\
\cite{Wang-10-Nov-A} & OWRN & DF & No & Freq. flat & Yes & No & N/A & Comp & Both & No & DSTC \\
\cite{Liu-14-Mar-A} & OWRN & DF & No & Freq. flat & Yes & No & N/A & Comp & Both & No & DLC-STC \\
\cite{Nasir-11-Jun-P} & OWRN & DF & No & Freq. flat & No & Yes & Blind & Both & Both & Yes & \\
\cite{Wang-12-Jun-P} & OWRN & DF & No & Freq. sel. & No & Yes & Pilot & Both & CFO & No & \\
\cite{Nasir-13-Jun-P} & OWRN & DF & No & Freq. flat & No & No & Pilot & Comp & Both & No & DSTBC \\
\cite{Chang-12-A} & TWRN & AF & No & Freq. sel. & No & Yes & Pilot & Both & TO & No & \\
\cite{Jiang-13-A} & TWRN & AF & No & Freq. flat & No & Yes & Pilot & Both & TO & Yes & \\
\cite{Zha-14-A} & TWRN & AF & No & Freq. flat & No & Yes & Semi-blind & Both & TO & No & \\
\cite{Wu-14-Jun-P} & TWRN & AF & No & Freq. flat & No & No & Pilot & Both & CFO & No & \\
\cite{Nasir-14-Jun-SPAWC-P} & TWRN & AF & No & Freq. flat & No & Yes & Pilot & Both & Both & Yes & \\
\cite{Ferrett-12-May-P}& TWRN & AF & No & Freq. flat & Yes & No & N/A & Comp & CFO & N/A & \\
\cite{Jain-11-Dec-P} & TWRN & DF & No & Freq. flat & No & Yes & Pilot & Both & TO & No & \\
\hline
 \end{tabular}
\label{tab:Coop_SC}
\vspace{1cm}
\caption{Summary of synchronization research in multi-carrier cooperative communication systems} \centering
\begin{tabular}{|l|l|l|l|l|l|l|l|l|l|l|l|} \hline
Article & Network & DF/AF & DL & Fading  & CSI Req. & CE &  Blind/Pilot & Est/Comp & TO/CFO  & Bound & Comments  \\ \hline
\cite{Jung-14-A} & OWRN & AF & No & Freq. sel. & No  & Yes & Pilot & Est & TO & No & Ranging method \\
\cite{Yao-12-A} & OWRN & AF & No & Freq. sel. & No & Yes & Pilot & Comp & CFO & No & \\
\cite{Zhang-12-Jun-A} & OWRN & AF & No & Freq. sel. & Yes & No & N/A & Comp & CFO & N/A & SFCC \\
\cite{Zhang-13-Mar-A} & OWRN & AF & No & Freq. sel. & Yes & No & N/A & Comp & CFO & N/A & Alamouti coding \\
\cite{Salim-14-Jun-P} & OWRN & AF & Yes & Freq. sel. & No & Yes & Pilot & Est & CFO & Yes & PHN \\
\cite{Zhang-12-May-P} & OWRN & Both & No & Freq. sel. & No & Yes & Pilot & Both & CFO & No & OSTBC \\
\cite{Won-13-Sep-P} & OWRN & AF & Yes & Freq. sel. & Yes & No & N/A & Comp & CFO & N/A & \\
\cite{Kim-11-Sep-A}& OWRN & DF & Yes & Freq. sel. & Yes & No & Pilot & Est & CFO & No & \\
\cite{Huang-10-J-A}& OWRN & DF & No & Freq. sel. & Yes & No & Pilot & Both & Both & No & \\
\cite{Guo-13-A}& OWRN & DF & No & Freq. sel. & Yes & No & Pilot & Est & Both & No & \\
\cite{Xiong-13-A}& OWRN & DF & No & Freq. sel. & Yes & No & Pilot & Both & CFO & No & SFBC \\
\cite{Huang-10-Nov-A}& OWRN & DF & No & Freq. sel. & Yes & No & N/A  & Comp & CFO & N/A & SFBC \\
\cite{Thiagarajan-10-Dec-P}& OWRN & DF & No & Freq. sel. & No & Yes & Pilot & Est & CFO & Yes & \\
\cite{Rugini-12-Jun-P}& OWRN & DF & No & Freq. sel. & No & Yes & Pilot & Est & CFO & No & \\
\cite{Wang-12-Dec-P}& OWRN & DF & No & Freq. sel. & Yes & No & Pilot & Both & CFO & No & Alamouti coding \\
\cite{Etezadi-10-Dec-P}& OWRN & DF & No & Freq. sel. & Yes & No & N/A & Comp & CFO & N/A & STC \\
\cite{Lu-10-Sep-P} & OWRN & DF & No & Freq. sel. & Yes & No & N/A & Comp & CFO & N/A & SFBC \\
\cite{Lu-11-May-P} & OWRN & DF & No & Freq. sel. & Yes & No & N/A & Comp & CFO & N/A & \\
\cite{Ponnaluri-10-Dec-P}& OWRN & DF & No & Freq. sel. & No & Yes & Pilot & Both & CFO & No & \\
\cite{Xiao-11-May-P}&  OWRN & DF & No & Freq. sel. & Yes & No & N/A & Comp & CFO & N/A & DLC-SFC \\
\cite{Yang-10-Dec-P}& OWRN & DF & No & Freq. sel. & Yes & No & Pilot & Est & Both & No & \\
\cite{Wang-13-Sep-P}& OWRN & DF & No & Freq. sel. & Yes & No & N/A & Comp & CFO & N/A & SFBC \\
\cite{Zhang-13-Sep-P}& OWRN & DF & No & Freq. sel. & Yes & No & Pilot & Est & Both & No & \\
\cite{Ho-13-A} & TWRN & AF & No & Freq. sel. & No & No & Pilot & Est & CFO & Yes & TS Design \\
\cite{Wang-11-A} & TWRN & AF & No & Freq. sel. & No & Yes & Pilot & Both & CFO & No & \\
\cite{Li-13-Dec-P} & TWRN & AF & No & Freq. sel. & Yes & No & N/A & Est & CFO & N/A & \\
\hline
 \end{tabular}
\label{tab:Coop_MC}
\end{table*}

In cooperative communication systems, the information transmission between the two communicating nodes is accomplished with the help of an intermediate relay. Let us assume a general scenario with the presence of multiple relays. There are two important types of cooperative communication networks:
\begin{itemize}
\item \textit{One-way relaying network (OWRN)}, where information transmission occurs in one direction via intermediate relays.

\item \textit{Two-way relaying network (TWRN)}, where information transmission occurs simultaneously in both directions and both nodes exchange their information with the help of intermediate relays.

\end{itemize}

The relays themselves can operate in different modes. The two most common modes are i) decode-and-forward (DF) and ii) amplify-and-forward (AF) operation. In DF mode, the relays decode the received signal and forward the decoded signal to the intended destination node(s). In AF mode, the relays do not decode the received message and simply amplify and forward the received signal.

In the following subsections, we review the recent literature that deals with timing and carrier synchronization in single-carrier and multi-carrier cooperative communication systems.

\subsection{Single-carrier cooperative communication systems}

Since the relaying operations are different in DF and AF cooperative communication systems, so to are their synchronization methodologies. In AF, for example, the relays may not be required to convert the received passband signal to baseband and to perform carrier synchronization \cite{Nasir-12-A}. Generally, the synchronization parameters to be estimated differ in OWRNs and TWRNs. In TWRN, there is self interference, which affects the way how the synchronization problem is formulated. In the following subsections, we provide separate literature reviews for synchronization in AF-OWRN, AF-TWRN, DF-OWRN, and DF-TWRN for single-carrier communication systems.

\subsubsection{Decode-and-Forward One-Way Relaying Network (DF-OWRN)}

\paragraph{System Model}\label{sec:DF-OWRN-SM-SC} The communication generally takes place in two phases. During the first phase, the source transmits the information to the relays. During the second phase, the relays decode the received signal and forward it to the destination. Typically, it is assumed that the direct communication link between the source and the destination is absent or blocked due to some obstacles. However, in general, there could be a direct communication link between them. In such a case, the destination also hears the source message during the first phase and coherently combines it with the message received during the second phase.

\paragraph{Synchronization Challenge}\label{sec:DF-OWRN-SC-SC} In DF-OWRN, during the first phase of the two-phase communication process, the synchronization between the source and the relays or between the source and the destination (in the presence of direct link) is achieved by estimating and compensating for a single TO and CFO between the source and each relay or between the source and the destination (in the presence of direct link). During the second phase, the synchronization between the relays and the destination is achieved by estimating and compensating for the multiple TOs and multiple CFOs between the multiple relays and the destination. Note that in the case of a single relay, only a single TO and a single CFO is required to be estimated and compensated for at the destination during the second communication phase. Increasing the number of relays raises the challenge of pilot design and estimation overhead.

\paragraph{Literature Review} The summary of the research carried out to achieve timing and carrier synchronization in single carrier DF-OWRN is given in Table \ref{tab:Coop_SC}:
\begin{enumerate}

\item Estimation or compensation of timing offsets alone and frequency offsets alone is studied in~\cite{Nasir-13-AT,Avram-12-Sep-P,Mehrpouyan-10-Dec-P} and~\cite{Wang-12-Jun-P,Wang-11-Feb-A,Liu-12-A,Mehr-11-A}, respectively.

\item Joint timing and carrier synchronization is studied in \cite{Nasir-11-Jun-P,Wang-10-Nov-A,Liu-14-Mar-A,Nasir-13-Jun-P,Nasir-12-A}.

\end{enumerate}

Further details or differences are provided in the last column of Table \ref{tab:Coop_SC}, which indicates whether STBC, SFBC, or training sequence (TS) design is considered.


\subsubsection{Decode-and-Forward Two-Way Relaying Network (DF-TWRN)}

\paragraph{System Model}\label{sec:DF-TWRN-SM-SC} TWRNs allow for more bandwidth efficient use of the available spectrum since they allow for simultaneous information exchange between the two nodes. In TWRNs, it is usually assumed that there is no direct communication link between the two nodes. During the first phase of the two-phase communication process, the information arrives at the relays from the two nodes. The signals from the two nodes are superimposed at the relays. During the second phase, the relays decode the exclusive OR (XOR) of the bits from the received superimposed signal and then broadcast a signal constructed from the XOR of the bits back to the two nodes \cite{Jain-11-Dec-P}.

\paragraph{Synchronization Challenge}\label{sec:DF-TWRN-SC-SC} The synchronization challenge during the first communication phase of DF-TWRN is unlike that for DF-OWRN. In DF-TWRN, the relays receive the superimposed signals from the two nodes during the first communication phase. Thus, unlike DF-OWRN, the received signal at each relay during the first communication phase is a function of two TOs and two CFOs, which need to be jointly estimated and compensated for in order to decode the modulo-2 sum of the bits from the two nodes. The synchronization challenge during the second communication phase of DF-TWRN is similar to that described for DF-OWRN in Section \ref{sec:DF-OWRN-SC-SC}. Another challenge for TWRN is the pilot design in the presence of self-interference at the relay node.

\paragraph{Literature Review} A summary of the research carried out to achieve timing and carrier synchronization in single carrier DF-TWRN is given in Table \ref{tab:Coop_SC}. There is only one paper in the last five years that falls in this category and proposes joint estimation and compensation of timing offsets \cite{Jain-11-Dec-P}. Most of the research has considered AF relaying for TWRN due to its implementation advantages.

\subsubsection{Amplify-and-Forward One-Way Relaying Network (AF-OWRN)}

\paragraph{System Model}\label{sec:AF-OWRN-SM-SC} The system model for AF-OWRN is similar to that of Section \ref{sec:DF-OWRN-SM-SC} for DF-OWRN. However, instead of DF operation, the relays amplify and forward the source information.

\paragraph{Synchronization Challenge}\label{sec:AF-OWRN-SC-SC} The synchronization challenge for AF-OWRN is quite similar to that described for DF-OWRN in Section \ref{sec:DF-OWRN-SM-SC}, except for one important difference. In DF-OWRN, the signal is decoded at the relay and the received signal at the destination is impaired by multiple TOs and multiple CFOs only between the relays and the destination. In AF-OWRN, the received signal at the destination is affected not only by the multiple TOs and multiple CFOs between the relays and the destination (as in the case of DF-OWRN), but also by the residual TOs and CFOs between the source and the relays. This is due to imperfect synchronization during the first communication phase and the amplified and forwarded signal from the relays is a function of the residual TOs and CFOs between the source and the respective relays.\footnote{Note that in a relay transceiver design, which is different from the conventional relay transceiver design described above, the authors have proposed to only establish timing synchronization at the relay and estimate and to compensate for the sum of multiple CFOs from the source-to-relays-to-destination at the destination~\cite{Nasir-12-A}.}

\paragraph{Literature Review} The summary of the research carried out to achieve timing and carrier synchronization in single carrier AF-OWRN is given in Table \ref{tab:Coop_SC}:
\begin{enumerate}
 \item Estimation or compensation of timing offset alone and frequency offset alone is studied in \cite{Li-10-A,Mehrpouyan-11-C-A,Mehrpouyan-10-Dec-P}
and \cite{Mehr-11-A}, respectively.

\item Joint timing and carrier synchronization is studied in~\cite{Nasir-13-A,Yadav-10-May-P,Nasir-12-A}.

\end{enumerate}

\subsubsection{Amplify-and-Forward Two-Way Relaying Network (AF-TWRN)}

\paragraph{System Model}\label{sec:AF-TWRN-SM-SC} During the first phase, similar to DF-TWRN, the relays receive the superimposed signal from the two nodes. During the second phase, the relays amplify and broadcast the superimposed signal back to the two nodes \cite{Jain-11-Dec-P}.

\paragraph{Synchronization Challenge}\label{sec:AF-TWRN-SC-SC} In AF-TWRN, when the relays receive the superimposed signals from the two nodes during the first communication phase, each relay only needs to carry out timing synchronization, i.e., estimate and compensate for the TOs between the two nodes and the relay \cite{Wu-14-Jun-P,Nasir-14-Jun-SPAWC-P}. The reason will be explained shortly. During the second communication phase, the relays amplify and broadcast the time-synchronized version of the superimposed signal. Each node then needs to estimate and compensate for the MTOs between the relays and itself and the sum of the multiple CFOs from the other node-to-relays-to-itself.

Note that each node in this case does not need to estimate and compensate for the multiple CFOs from itself to relays to the other node because the effect of CFOs between itself and the relays during the first communication phase is canceled by the effect of CFOs between the relays and itself during the second communication phase due to the use of the same oscillators \cite{Wu-14-Jun-P,Nasir-14-Jun-SPAWC-P}. Due to this very reason, the authors in \cite{Wu-14-Jun-P,Nasir-14-Jun-SPAWC-P} propose to only perform timing synchronization at the relay nodes during the first communication phase in AF-TWRN.

\paragraph{Literature Review} The summary of the research carried out to achieve timing and carrier synchronization in single carrier AF-OWRN is given in Table \ref{tab:Coop_SC}:
 \begin{enumerate}

 \item The estimation or compensation of timing offset alone and frequency offset alone is studied by \cite{Chang-12-A,Jiang-13-A,Zha-14-A}
and \cite{Wu-14-Jun-P,Ferrett-12-May-P}, respectively.

\item The joint timing and carrier synchronization is studied by \cite{Nasir-14-Jun-SPAWC-P}.

 \end{enumerate}

\subsubsection{Summary} In single-carrier cooperative communication systems, synchronization in the presence of the direct link is an important research problem, which has not been considered except in a few papers. Also optimum training sequence design is an interesting open research problem.

\subsection{Multi-carrier cooperative communication systems}

The following subsections review the literature for synchronization in AF-OWRN, AF-TWRN, DF-OWRN, and DF-TWRN for multi-carrier communication systems. Since, most of the papers consider orthogonal frequency division multiplexing (OFDM) as a special case of multi-carrier communication system, the system model and synchronization challenge below is presented for OFDM systems.

\subsubsection{Decode-and-Forward One-Way Relaying Network (DF-OWRN)}

\paragraph{System Model} \label{sec:DF-OWRN-SM-MC} Apart from the IFFT/CP addition and CP removal/FFT operations at the transmitter and receiver side at each node, respectively, the system model for DF-OWRN for multicarrier systems is similar to that described for single carrier DF-OWRN presented in Section \ref{sec:DF-OWRN-SM-SC}.

\paragraph{Synchronization Challenge}  \label{sec:DF-OWRN-SC-MC} The synchronization challenge during the first communication phase between the source and the relays is similar to that presented for SISO multi-carrier systems in Section \ref{sec:SISO-MC-SC}. During the second communication phase, the relays decode the received signal and forward it to the destination. Thus, the resulting signal at the destination is affected by multiple TOs and multiple CFOs resulting in strong ISI and ICI. The synchronization challenge is to jointly estimate and compensate for the effect of multiple TOs and multiple CFOs in order to mitigate ISI and ICI.

\paragraph{Literature Review} The summary of the research carried out to achieve timing and carrier synchronization in single carrier DF-OWRN is given in Table \ref{tab:Coop_MC}:
\begin{enumerate}

\item The estimation or compensation of frequency offset alone is studied in \cite{Zhang-12-May-P,Kim-11-Sep-A,Xiong-13-A,
Huang-10-Nov-A,Thiagarajan-10-Dec-P,Rugini-12-Jun-P,Wang-12-Dec-P,Etezadi-10-Dec-P,Lu-10-Sep-P,
Lu-11-May-P,Ponnaluri-10-Dec-P,Xiao-11-May-P,Wang-13-Sep-P}.

\item The joint timing and carrier synchronization is studied in \cite{Huang-10-J-A,Guo-13-A,Yang-10-Dec-P,Zhang-13-Sep-P}.

\end{enumerate}

Further details or differences are provided in the last column of Table \ref{tab:Coop_MC}, which indicates if STBC or SFBC is considered.


\subsubsection{Decode-and-Forward Two-Way Relaying Network (DF-TWRN)}

\paragraph{System Model} Other than the IFFT/CP addition and CP removal/FFT operations at the transmitter and receiver side at each node, respectively, the system model for DF-TWRN for multicarrier systems is similar to that for single carrier systems presented in Section \ref{sec:DF-TWRN-SM-SC}.

\paragraph{Synchronization Challenge} The received signal at each relay during the first communication phase is a function of two TOs and two CFOs, which need to be jointly estimated and compensated. Thus, the synchronization challenge is similar to that described for the second communication phase of DF-OWRN with two TOs and two CFOs. Moreover, the synchronization challenge during the second communication phase of DF-TWRN is also similar to the one described for the second communication phase of DF-OWRN in Section \ref{sec:DF-OWRN-SC-MC}.

\paragraph{Literature Review} To the best of our knowledge, no paper in the last five years falls into this category, since the research in the synchronization of multi-carrier TWRN has considered AF relaying.

\subsubsection{Amplify-and-Forward One-Way Relaying Network (AF-OWRN)}

\paragraph{System Model} The system model for AF-OWRN for multicarrier systems is similar to the one for single carrier systems presented in Section \ref{sec:AF-OWRN-SM-SC}. However, being an OFDM system, there are IFFT/CP addition and CP removal/FFT operations at the source transmitter and the destination receiver, respectively.  Note that unlike DF-OWRN for multicarrier systems, the FFT and IFFT operations are not usually conducted at the receiver and transmitter of the relays, respectively.

\paragraph{Synchronization Challenge} The synchronization challenge during the first communication phase for AF-OWRN is similar to the one described for DF-OWRN in Section \ref{sec:DF-OWRN-SC-MC}. During the second communication phase, similar to AF-OWRN in single-carrier systems, the received signal at the destination is affected by not only the multiple TOs and multiple CFOs between the relays and the destination, but also by the residual TOs and CFOs between the source and the relays, which can cause additional ISI and ICI. The effect of ISI and ICI is required to be mitigated to achieve synchronization.

\paragraph{Literature Review} The summary of the research carried out to achieve timing and carrier synchronization in single carrier AF-OWRN is given in Table \ref{tab:Coop_MC}. The estimation or compensation of timing offset alone and frequency offset alone is studied in \cite{Won-13-Sep-P,
Zhang-12-Jun-A,Zhang-13-Mar-A,Salim-14-Jun-P,Zhang-12-May-P,Yao-12-A} and \cite{Jung-14-A}, respectively. Further details or differences are provided in the last column of Table \ref{tab:Coop_MC}, which indicates if any additional parameter, e.g., phase noise (PHN) estimation is considered or if STBC or any other type of space coding is considered.

\subsubsection{Amplify-and-Forward Two-Way Relaying Network (AF-TWRN)}

\paragraph{System Model} Apart from the IFFT/CP addition and CP removal/FFT operations at the transmitter and receiver, respectively, the system model for AF-TWRN for multicarrier systems is similar to the one for single carrier systems presented in Section \ref{sec:AF-TWRN-SM-SC}. Note that unlike DF-TWRN for multicarrier systems, the FFT and IFFT operations are not usually conducted at the receiver and transmitter of the relays, respectively.

\paragraph{Synchronization Challenge} The required parameters to be estimated and compensated to achieve synchronization in multi-carrier AF-TWRN are similar to the one described in the synchronization challenge of single-carrier AF-TWRN in Section \ref{sec:AF-TWRN-SM-SC}. The difference is to mitigate the effect of ISI and ICI in OFDM systems.

\paragraph{Literature Review} The summary of the research carried out to achieve timing and carrier synchronization in single carrier AF-OWRN is given in Table \ref{tab:Coop_MC}. The estimation or compensation of frequency offset alone is studied in \cite{Ho-13-A,Wang-11-A,Li-13-Dec-P}. The categorized papers differ in a sense that training sequence design is proposed in \cite{Ho-13-A}, joint CFO estimation and compensation with channel estimation is studied in \cite{Wang-11-A}, and CFO estimation alone is proposed in \cite{Li-13-Dec-P}.

\subsubsection{Summary} In multi-carrier cooperative communication systems, which has been the subject of intense research in the past five years, the majority of the papers have considered decode-and-forward for one-way relaying network and amplify-and-forward for two-way relaying network. In addition, most of the solutions are pilot based. Hence, it is a challenging open problem to design semi-blind and blind estimators.

\section{Multiuser/Multicell Interference Networks}\label{sec:interference}

\subsection{SC-FDMA uplink communication systems}

Single-carrier frequency division multiple access (SC-FDMA) is an extension of SC-FDE to accommodate multiple users. In SC-FDMA uplink communication systems, multiple users communicate with a single receiver and the effect of channel distortion is equalized in frequency domain at the receiver. Like the SC-FDE receiver, FFT and IFFT modules are present in the SC-FDMA receiver. However, the SC-FDMA transmitter also incorporates FFT and IFFT modules. Disjoint sets of $M$ subcarriers are assigned to each of the $K$ users and data symbols from each user are modulated over a unique set of subcarriers through an $M$-point FFT operation. Next, $KM$-point IFFT is applied to transform the signal to time domain, such that the output of FFT is applied to the user specified $M$ inputs of IFFT block and $0$ is applied to the remaining $(K-1)M$ inputs of IFFT block. Next, cyclic prefix is appended at the start of the transmission block to mitigate the multipath channel effect. At the receiver side, following $KM$-point FFT and frequency domain equalization, $K$ of $M$-point IFFT operations are applied to decode the information from the $K$ users. SC-FDMA can also be interpreted as a linearly precoded orthogonal frequency division multiple access (OFDMA) scheme, in the sense that it has an additional FFT processing step preceding the conventional OFDMA processing.

Due to the presence of an independent oscillator at each transmitting user and due to different propagation delays between each user and the receiver, the received signal in the SC-FDMA uplink communication system suffers from multiple CFOs and multiple TOs. The combined effect of TOs and CFOs results in both ISI and loss of orthogonality among subcarriers, which, in turn, generates ICI and multiple access interference (MAI) at the receiver. Thus, in order to decode information from each user, a receiver has to estimate multiple TOs and multiple CFOs to compensate the effect of ICI, ISI, and MAI. In uplink communication systems, there is another synchronization challenge that the correction of one user's frequency and timing at the receiver can misalign those of the other users \cite{Zhang-Ryu-10-May-A}.

It must be noted that long term evolution (LTE) systems~\cite{Sesia-B-09} use SC-FDMA for communication in the uplink. Synchronization is achieved through periodically transmitted primary and secondary synchronization signals from the base station. Any user who has not yet acquired the uplink synchronization can use the primary and secondary synchronization signals (transmitted by the base station) to first achieve synchronization in the downlink. Next, compensation for the propagation loss is made as part of the uplink random access procedure~\cite{Sesia-B-09}.

The research carried out to achieve timing and carrier synchronization in SC-FDMA uplink communication systems is summarized in Table \ref{tab:SC-FDMA}:
\begin{enumerate}

\item The estimation or compensation of multiple CFOs alone is studied in \cite{Song-11-A,
Meng-10-A,Kamali-12-A,Zhu-10-A,Zhang-Ryu-10-May-A,Iqbal-14-A,Kamali-14-A,Kamali-11-Mar-A,Fu-14-A,Chen-10-May-P}. Though \cite{Song-11-A,Meng-10-A,Kamali-12-A,Kamali-14-A,Kamali-11-Mar-A,Fu-14-A,Chen-10-May-P} consider the same channel model with known CSI in order to propose multiple CFO compensation, they differ in the following characteristics. Different channel allocation strategies and their effects on interference due to CFO is studied in \cite{Song-11-A}. An interference self-cancellation scheme to compensate for the effect of CFO is proposed by \cite{Meng-10-A}.

\item Joint timing and carrier synchronization is studied in \cite{Darsena-13-A}. Further, \cite{Kamali-12-A} and \cite{Fu-14-A} consider multiple antennas at the users and receiver, \cite{Fu-14-A} proposes blind beamforming assuming that the multipath delay is greater than the cyclic prefix length, and \cite{Chen-10-May-P} proposes MMSE-FDE. Finally, joint timing and carrier synchronization in SC-FDMA systems is proposed in \cite{Darsena-13-A}.

\end{enumerate}

\subsection{OFDMA uplink communication systems}


An OFDMA communication system is an extension of an OFDM system to accommodate multiple users.  In OFDMA uplink communication systems, multiple users communicate with a single receiver. Unlike OFDM systems, where information of a single user is modulated over all subcarriers, in OFDMA uplink transmitter, each user transmits over a set of assigned subcarriers. For each user, group of $M$ modulated symbols are applied to the user specified $M$ inputs of IFFT block and $0$ is applied to the remaining $(K-1)M$ inputs of IFFT block. In the end, cyclic prefix is appended and information is transmitted from every user. At the receiver side, following cyclic prefix removal and FFT operation, equalization is carried out to decode the information from the $K$ users.

Similar to SC-FDMA uplink communication system, the received signal in SC-FDMA uplink suffers from multiple CFOs and multiple TOs. The combined effects of TOs and CFOs results in ISI, ICI, and MAI. The receiver has to estimate and compensate the effect of multiple TOs and multiple CFOs in order to decode information from each user. Similar to SC-FDMA uplink communication systems, the correction of one user's frequency and timing at the receiver can misalign the other users.

The research to achieve timing and carrier synchronization in OFDMA uplink communication systems is summarized in Table \ref{tab:OFDMA-UL}. All categorized papers study carrier synchronization only \cite{Kim-10-A,Pengfei-A-10,Shah-13-A,Ho-14-A,Hsieh-11-A,Zhang-14-M-A,Zhang-12-A,Lee-12-A,Keum-14-A,Nguyen-14-A,Estrella-13-A,Bai-12-Jun-A,Muneer-14-A,Lee-12-Aug-A,Peng-12-A,Fa-13-A,Hou-12-A,Yerramalli-10-Dec-P,Aziz-11-Dec-P,Chen-10-Dec-P,Bertrand-10-May-P,Chang-10-May-P,Wu-10-May-P,Zhang-10-May-VTC-P,Xiong-12-Sep-P,Tu-12-May-P,Farhang-13-Jun-P}. Though they consider the same channel model, they differ in respect of providing joint channel estimation, requiring pilot/training, proposing estimation or compensation of multiple CFO, or providing lower bound. Note that further details or differences are provided in the last column of Table \ref{tab:OFDMA-UL}.

\subsection{CDMA communication systems} \label{sec:CDMA}
\begin{table*}[p]
\vspace{1cm}
\caption{Summary of synchronization research in SC-FDMA uplink communication systems} \centering
\begin{tabular}{|l|l|l|l|l|l|l|l|l|} \hline
Article & Fading  & CSI Req. & CE &  Blind/Pilot & Est/Comp & TO/CFO  & Bound & Comments  \\ \hline
\cite{Darsena-13-A} & Freq. sel. & No & Yes & Blind & Both & Both & No & \\
\cite{Song-11-A} & Freq. sel. & Yes & No & N/A & Comp & CFO & N/A & Channel Allocation, OFDMA \\
\cite{Meng-10-A} & Freq. sel. & Yes & No & N/A & Comp & CFO & N/A & Interference Cancelation \\
\cite{Kamali-12-A} & Freq. sel. & No & No & N/A & Comp & CFO & N/A & MIMO \\
\cite{Zhu-10-A} & Freq. sel. & No & No & Blind & Both & CFO & N/A &  \\
\cite{Zhang-Ryu-10-May-A} & Freq. sel. & Yes & No & Pilot & Both & CFO & No &  \\
\cite{Iqbal-14-A} & Freq. sel. & No & Not req. & Blind & Comp & CFO & No & MIMO AFD-DFE \\
\cite{Kamali-14-A} & Freq. sel. & No & No & N/A & Comp & CFO & N/A & \\
\cite{Kamali-11-Mar-A}& Freq. sel. & No & No & N/A & Comp & CFO & N/A & \\
\cite{Fu-14-A} & Freq. sel. & No & No & N/A & Comp & CFO & N/A & Multi-antenna users and blind beamforming \\
\cite{Chen-10-May-P} & Freq. sel. & No & No & N/A & Comp & CFO & N/A & MMSE-FDE \\
\hline
 \end{tabular}
\label{tab:SC-FDMA}
\vspace{1cm}
\caption{Summary of synchronization research in OFDMA uplink communication systems} \centering
\begin{tabular}{|l|l|l|l|l|l|l|l|l|} \hline
Article & Fading  & CSI Req. & CE &  Blind/Pilot & Est/Comp & TO/CFO  & Bound & Comments  \\ \hline
\cite{Kim-10-A}    & Freq. sel. & Yes & No  & Pilot & Both & CFO & No & MIMO, Rao-Blackwellized particle filter \\
\cite{Pengfei-A-10}& Freq. sel. & Yes & No  & Pilot & Both & CFO & No & CFO tracking \\
\cite{Shah-13-A}   & Freq. sel. & Yes & No  & Pilot & Est  & CFO & No & canonical particle swarm optimisation scheme \\
\cite{Ho-14-A}     & Freq. sel. & Yes & No  & Pilot & Est  & CFO & Yes& alternating projection method \\
\cite{Hsieh-11-A}  & Freq. sel. & Yes & No  & Blind & Est  & CFO & Yes& Interleaved OFDMA \\
\cite{Zhang-14-M-A}& Freq. sel. & Yes & No  & Blind & Both & CFO & No & tile-based OFDMA \\
\cite{Zhang-12-A}  & Freq. sel. & Yes & Yes & Blind & Est  & CFO & No & Interleaved OFDMA \\
\cite{Lee-12-A}    & Freq. sel. & Yes & No  & Pilot & Both & CFO & Yes& conjugate gradient method \\
\cite{Keum-14-A}   & Freq. sel. & Yes & No  & Pilot & Est  & CFO & Yes& Interleaved OFDMA \\
\cite{Nguyen-14-A} & Freq. sel. & Yes & No  & Pilot & Both & CFO & No & multiuser interference cancellation based algorithm \\
\cite{Estrella-13-A}& Freq. sel.& Yes & No  & Blind & Est  & CFO & No & LS estimation \\
\cite{Morelli-10-A}& Freq. sel. & Yes & No  & Pilot & Est  & CFO & Yes& SFO estimation \\
\cite{Movahhedian-10-A}& Freq. sel.& Yes& No& Blind & Est  & CFO & Yes& precoded OFDMA \\
\cite{Sourck-11-A} & Freq. sel. & Yes & No  & Pilot & Both & CFO & No & OFDM and FBMC \\
\cite{Song-11-A}   & Freq. sel. & Yes & No  & N/A   & Comp & CFO & N/A& Channel Allocation, SC-FDMA \\
\cite{Haring-10-A} & Freq. sel. & Yes & Yes & Pilot & Est  & CFO & Yes& iterative ML est \\
\cite{Miao-10-A}   & Freq. sel. & Yes & No  & Pilot & Est  & CFO & No & Ranging method \\
\cite{Bai-12-Jun-A}& Freq. sel. & Yes & No  & Pilot & Est  & CFO & Yes& tile-based OFDMA \\
\cite{Muneer-14-A} & Freq. sel. & Yes & Yes & Pilot & Both & CFO & Yes& doubly selective channel \\
\cite{Lee-12-Aug-A}& Freq. sel. & Yes & No  & Pilot & Est  & CFO & Yes& conjugate gradient method \\
\cite{Peng-12-A}   & Freq. sel. & Yes & No  & N/A   & Comp & CFO & N/A& Interleaved OFDMA \\
\cite{Fa-13-A}     & Freq. sel. & Yes & No  & N/A   & Comp & CFO & N/A& multistage interference cancellation \\
\cite{Hou-12-A}    & Freq. sel. & Yes & No  & N/A   & Comp & CFO & N/A& successive interference cancellation \\
\cite{Yerramalli-10-Dec-P}  & Freq. sel. & Yes & Yes  & Pilot   & Est & CFO & No& partial FFT modulation \\
\cite{Aziz-11-Dec-P}& Freq. sel.& Yes & No  & N/A   & Comp & CFO & N/A& subcarrier allocation algorithm \\
\cite{Chen-10-Dec-P}& Freq. sel.& Yes & No  & Pilot & Est  & CFO & Yes& MIMO OFDMA, Bayesian Estimation \\
\cite{Bertrand-10-May-P}& Freq. sel. & Yes & Yes & Pilot & Both & CFO & No& high mobility \\
\cite{Chang-10-May-P}   & Freq. sel. & Yes & Yes & Pilot & Both & CFO & No& common CFO estimation\\
\cite{Wu-10-May-P}      & Freq. sel. & Yes & Yes & Blind & Est  & CFO & No& interleaved OFDMA, DoA estimation\\
\cite{Zhang-10-May-VTC-P}&Freq. sel. & Yes & No  & N/A   & Comp & CFO & N/A& exploit time domain inverse matrix \\
\cite{Xiong-12-Sep-P}   & Freq. sel. & Yes & No  & Pilot & Both & CFO & No& ad hoc network\\
\cite{Tu-12-May-P}      & Freq. sel. & Yes & No  & N/A   & Comp & CFO & N/A& soft interference cancellation \\
\cite{Farhang-13-Jun-P} & Freq. sel. & Yes & No  & N/A   & Comp & CFO & N/A& LS and MMSE based compensation \\
\hline
 \end{tabular}
\label{tab:OFDMA-UL}
\vspace{1cm}
\caption{Summary of synchronization research in CDMA communication systems} \centering
\begin{tabular}{|l|l|l|l|l|l|l|l|l|l|} \hline
Article & SC/MC & Fading  & CSI Req. & CE &  Blind/Pilot & Est/Comp & TO/CFO  & Bound & Comments  \\ \hline
\cite{Xu-13-A} & SC & Freq. flat & Yes & No & Blind & Est & CFO & No & \\
\cite{Wang-Coon-11-A} & SC & Freq. sel. & Yes & No & N/A & Comp & CFO & N/A & Spreading code design \\
\cite{Bangwon-12-A} & MC & Freq. sel. & Yes & No & Pilot & Comp & CFO & No & Multiuser detector \\ 
\cite{Tadjpour-10-A} & MC & Freq. sel. & Yes & No & N/A & Comp & CFO & N/A & Multiuser detector \\
\cite{Lin-12-Oct-A} & MC & Freq. sel. & No & Yes & Blind & Both & CFO & No & STBC MIMO \\
\cite{Manglani-11-A} & MC & Freq. sel. & Yes & No & Pilot & Comp & CFO & No & \\
\cite{Li-10-Dec-P} & MC & Freq.sel. & Yes & No & Blind & Both & CFO & No & \\
\cite{Yan-14-Art} & MC & Freq. sel. & No & Yes & Pilot & Comp & CFO &  N/A & \\
\hline
 \end{tabular}
\label{tab:CDMA-MU}
\end{table*}

Code division multiple access (CDMA) communication systems allow multiple transmitters to send information simultaneously to a single receiver. All users share the same frequency and time resources. To permit this to be achieved without undue interference, CDMA employs spread-spectrum technology, i.e., either direct-sequence spread-spectrum (DS-CDMA) or frequency hopping spread spectrum (FH-CDMA):
\begin{itemize}

\item In DS-CDMA, a unique (orthogonal or nearly orthogonal) spreading code is assigned to each transmitter. A modulated signal for each user is spread with a unique spreading code at the transmitter. Finally, at the receiver, information for each user is desperead by using the same unique despreading code. All users share the full available spectrum. Note that multi-carrier CDMA (MC-CDMA) spreads each user signal in the frequency domain, i.e., each user signal is carried over multiple parallel subcarriers and the spreading codes differ per subcarrier and per user.

\item FH-CDMA uses a short-term assignment of a frequency band to various signal sources. At each successive time slot of brief duration, the frequency band assignments are reordered. Each user employs a pseudo noise (PN) code, orthogonal (or nearly orthogonal) to all the other user codes, that dictates the frequency hopping band assignments.

\end{itemize}

Due to MUI, timing and carrier synchronization in CDMA is very challenging since compensating TO and CFO for one user may cause synchronization loss to other users. Particularly in CDMA systems, the presence of TOs and CFOs destroys the orthogonality of the spreading codes. The design of special spreading and despreading codes that are robust to synchronization errors is also a challenging task. Moreover, since the received signal powers from different users may vary significantly, the correct synchronized despreading may be challenging. Synchronization also faces multipath delay spread since PN chips have a short time duration. In MC-CDMA systems, due to the presence of CFOs, orthogonality among the subcarriers is lost and ICI is generated. Thus, the synchronization challenge is to estimate multiple TOs and multiple CFOs, compensate for MUI and ICI, and design proper spreading and despreading codes to decode information from each user.

Compared to DS-CDMA, the synchronization challenge is slightly reduced in FH-CDMA, because FH-CDMA synchronization has to be within a fraction of a hop time. Since spectral spreading does not use a very high hopping frequency but rather a large hop-set, the hop time will be much longer than the DS-CDMA system chip time. Thus, an FH-CDMA system allows for a larger synchronization error \cite{Chen-2007}. Moreover, in FH-CDMA, there is no need for synchronization among user groups, only between transmitter and receiver within a group is required.

Table \ref{tab:CDMA-MU} summarizes the research carried out to achieve timing and carrier synchronization in CDMA multiuser communication systems. All categorized papers consider direct-sequence (DS) spread spectrum technology and study the carrier synchronization alone \cite{Xu-13-A,Wang-Coon-11-A,Bangwon-12-A,Tadjpour-10-A,Lin-12-Oct-A,Manglani-11-A,Li-10-Dec-P,Yan-14-Art}, where as identified in the second column of Table \ref{tab:CDMA-MU}, \cite{Xu-13-A,Wang-Coon-11-A} consider single carrier communication and \cite{Bangwon-12-A,Tadjpour-10-A,Lin-12-Oct-A,Manglani-11-A,Li-10-Dec-P,Yan-14-Art} consider multi-carrier communication. In addition, as detailed in Table \ref{tab:CDMA-MU}, the categorized papers differ in respect of channel model, providing joint channel estimation, requiring pilot/training, proposing estimation or compensation of multiple CFO, or providing lower bound. Note that further details or differences are provided in the last column of Table \ref{tab:CDMA-MU}.

\subsection{Cognitive radio based communication systems}\label{sec:CR}

Cognitive radio networks allow unlicensed secondary users (SUs) access to the spectrum of the licensed primary users (PUs), without impairing the performance of the PUs. Depending on the spectrum access strategy, there are three main cognitive radio network paradigms~\cite{Goldsmith-2009}:
\begin{itemize}

\item In the \textit{underlay cognitive networks}, SUs can concurrently use the spectrum occupied by a PU by guaranteeing that the interference at the PU is below some acceptable threshold~\cite{Jing-2015}. Thus, SUs must know the channel strengths to the PUs and are also allowed to communicate with each other in order to sense how much interference is being created to the PUs.

\item In the \textit{overlay cognitive networks}, there is tight interaction and active cooperation between the PUs and the SUs. Thus, SUs use sophisticated signal processing and coding to maintain or improve the PU transmissions while also obtaining some additional bandwidth for their own transmission.

\item In \textit{interweave cognitive networks}, the SUs are not allowed to cause any interference to the PUs. Thus, SUs must periodically sense the environment to detect spectrum occupancy and transmit opportunistically only when the PUs are silent~\cite{Goldsmith-2009}.

\end{itemize}

In the context of \textit{interweave cognitive networks}, secondary users (SUs) sense the spectrum to detect the presence or absence of PUs, and use the unoccupied bands while maintaining a predefined probability of missed detection. Different methods are used to detect the presence of PUs such as matched filtering, energy detection, cyclostationary detection, wavelet detection and covariance detection. For multi-carrier OFDM systems, PUs adopt OFDM modulation and SUs make use of cyclic prefix, pilot tones of OFDM signals or cyclostationarity to detect the PUs. In cognitive radio based communication systems, the presence of timing and frequency offset affects the spectrum sensing performance and may result in false detection by the SUs. Accurate estimation and compensation of timing and frequency offsets from the PUs' signal is necessary to lead to the correct decision about the spectrum availability. However, this is challenging given that SUs do not have access to pilot symbols and may need to estimate these parameters in a blind fashion. Blind synchronization algorithms are also known to be less accurate, which introduces new challenges to spectrum sensing in the presence of synchronization errors in cognitive radio networks.

Recent research in synchronization for cognitive radio based communication systems is summarized in Table~\ref{tab:CR}. The papers in Table~\ref{tab:CR} assume interweave cognitive networks, except for~\cite{Xu-10-A} and~\cite{Zhou-11-Dec-P}, which consider underlay and both underlay/overlay based cognitive networks. In Table~\ref{tab:CR}:
\begin{enumerate}

\item The carrier synchronization alone is studied in \cite{Rebeiz-13-A,Nevat-12-Apr-P,
Xu-10-A,Zeng-13-A,Ding-13-Dec-P,Zhao-11-Dec-P,Zivkovic-11-Dec-P,Zhou-11-Dec-P}, where as identified in second column of Table \ref{tab:CR}, \cite{Rebeiz-13-A,Nevat-12-Apr-P} consider single carrier communication and \cite{Xu-10-A,Zeng-13-A,Ding-13-Dec-P,Zhao-11-Dec-P,Zivkovic-11-Dec-P,Zhou-11-Dec-P} consider multi-carrier communication.

\item The joint timing and carrier synchronization is proposed in \cite{Axell-12-A,Chen-12-Jan-A,Al-Habashna-12-A,Cheraghi-12-A,Liu-12-May-P}, where as identified in second column of Table \ref{tab:CR}, \cite{Axell-12-A} considers single carrier communication and \cite{Chen-12-Jan-A,Al-Habashna-12-A,Cheraghi-12-A,Liu-12-May-P} consider multi-carrier communication.

\end{enumerate}
The number of primary users are listed under the ``PUs" column. As detailed in Table \ref{tab:CR}, the categorized papers differ in channel model, providing joint channel estimation, requiring CSI or training, proposing estimation or compensation, or providing lower bound. If applicable, further details for some paper, e.g,. training sequence design, STBC, MIMO, or hardware design consideration, are provided in the last column of Table \ref{tab:CR}.

\subsection{Distributed multiuser communication systems}

In distributed multiuser communication systems, multiple distributed users try to communicate with a common receiver. Cooperation may exist among the distributed users to transmit the same information, i.e., broadcasting. On the other hand, each user might send its own data, which can cause MUI at the receiver. The receiver can employ successive interference cancelation to decode information from the desired user. Both single carrier and multi-carrier modulation schemes can be employed by the multiple users. Due to the presence of independent oscillator at each transmitting user, the Doppler effect, and the existence of a different propagation delay between each user and the receiver, the received signal may suffer from multiple CFOs and multiple TOs. The receiver has to jointly estimate and compensate the effect of these synchronization impairments in order to decode the desired information. Note that users in distributed multiuser communication systems may cooperate amongst themselves, which may not be possible in cooperative relaying networks due to the latter's simpler assumed relay topology.

Table \ref{tab:DMU} summarizes the research carried out to achieve timing and carrier synchronization in distributed multiuser communication systems. All listed papers \cite{Liu-12-Aug-A,Caus-11-Jun-P,
Zhi-11-May-P,
Sanchez-11-Jun-P} consider multi-carrier communication. Carrier synchronization alone is studied by \cite{Liu-12-Aug-A}. The joint timing and carrier synchronization is proposed in \cite{Caus-11-Jun-P,
Zhi-11-May-P,
Sanchez-11-Jun-P}, where joint estimation and compensation design is considered by \cite{Zhi-11-May-P} and only compensation of timing and carrier frequency offsets is studied in~\cite{Caus-11-Jun-P,Sanchez-11-Jun-P}. Different from \cite{Caus-11-Jun-P}, \cite{Sanchez-11-Jun-P} also considers joint channel estimation.

\subsection{CoMP based communication systems}

In coordinated multipoint transmission/reception (CoMP) communication systems, geographically separated base stations (also referred to as transmission points) coordinate and jointly process the signal transmission to multiple users at the cell edges. Thus, CoMP is also referred to as multicell cooperation. CoMP techniques can be broadly classified into i) coordinated scheduling and coordinated beamforming (CS/CB), ii) joint transmission (JT), and iii) transmission point selection (TPS) \cite{Lee-12-Feb-A}. Note that the transmission from the base stations can take place over single or multiple carriers. Some main points regarding the CoMP techniques are as follows:
\begin{itemize}
\item In \textit{coordinated scheduling and coordinated beamforming}, multiple coordinated transmission points only share the CSI for multiple users. The data packets that need to be conveyed to the users are available only at the respective transmission point to which each user belongs. Thus every base station coordinates with its cell edge user through beamforming. The users may experience residual interference from the other cells' communication depending on the location of the cell edge users.
\item In \textit{joint transmission}, multiple coordinated transmission points share both CSI and the data packets to be conveyed to all users. Thus, the same data is simultaneously transmitted to the intended user from multiple coordinated transmission points with appropriate beamforming weights.
\item \textit{Transmission point selection} can be regarded as a special form of JT, where transmission of beamformed data for a given user is performed at a single transmission point at each time instance. In addition, both CSI and the data is assumed to be available at multiple coordinated transmission points. Thus, an appropriate transmission point with access to the best channel conditions for individual users can be scheduled. While one transmission point coordinates with the scheduled user, other transmission points may possibly communicate in parallel to their respective users.
\end{itemize}

The synchronization between coordinating base stations and the users can be achieved either in uplink or downlink transmission. For \textit{downlink CoMP}, due to the different oscillators at the base stations and the different propagation delays between each base station and the user, the received signal at the user suffers from multiple CFOs and multiple TOs. The receiver has to estimate these parameters jointly and compensate for their effects in order to establish successful CB, JT, or TPS scheme among coordinating base stations. For \textit{uplink CoMP}, multiple users communicate with the base stations, causing multiple CFOs and multiple TOs at the base station. The base stations have to compensate for the effect of these multiple synchronization parameters in order to synchronize data transmission to the users during downlink communication by adopting any CoMP scheme. Since CoMP uses the backhaul link for coordination among the base stations, synchronization parameters can be exchanged in order to enhance the synchronization performance.

Considering synchronization in CoMP based communication systems, Table \ref{tab:CoMP} summarizes the recent research:
\begin{enumerate}

\item The timing synchronization alone is studied by \cite{Silva-14-A,Zhao-14-A,Iwelski-14-A}, where as identified in the second column of Table \ref{tab:CoMP}, \cite{Silva-14-A} considers single carrier communication and \cite{Zhao-14-A,Iwelski-14-A} consider multi-carrier communication.

\item Carrier synchronization alone is studied by \cite{Zarikoff-10-A,Liang-12-A,Jiang-14-A,Tsai-13-A,Vakilian-13-Dec-P,Pec-14-A}, where as identified in the second column of Table \ref{tab:CoMP}, \cite{Zarikoff-10-A} considers single carrier communication and \cite{Liang-12-A,Jiang-14-A,Tsai-13-A,Vakilian-13-Dec-P,Pec-14-A} consider multi-carrier communication.

\item Joint timing and carrier synchronization considering single and multi-carrier communication is analyzed in \cite{Liu-13-Apr-P} and \cite{Koivisto-13-Sep-P}, respectively.

\end{enumerate}

The type of CoMP communication, CB, JT, or TPS, is indicated under the ``Type" field in Table \ref{tab:CoMP}, where a blank field indicates that either only the training/estimation phase or uplink communication is considered by the authors. As presented in Table \ref{tab:CoMP}, the categorized papers differ based on assuming varying channel models, considering channel estimation, assuming the need for CSI or training, proposing either estimation or detection algorithms in the presence of synchronization errors, or driving lower bounds on estimation of TO and CFO. Further details about the assumed channel models in each paper are provided in the last column of Table \ref{tab:CoMP}.
\begin{table*}[p]
\vspace{1cm}
\caption{Summary of synchronization research in Cognitive Radio based communication systems} \centering
\begin{tabular}{|l|l|l|l|l|l|l|l|l|l|l|} \hline
Article & SC/MC & Channel Model & CSI Req. & CE &  Blind/Pilot & Est/Comp & TO/CFO  & Bound & PUs & Comments  \\ \hline
\cite{Axell-12-A} & SC & Freq. flat & No & No & N/A & Comp & Both & N/A & 1 & OSTBC MIMO \\
\cite{Rebeiz-13-A} & SC & AWGN & N/A & N/A & Blind & Comp & CFO & No & $>1$ & SCO \\
\cite{Nevat-12-Apr-P} & SC & Freq. flat & No & No & Blind & Both & CFO & No & 1 & \\
\cite{Xu-10-A} &  MC & Freq. sel. & Yes & No & N/A & Est & CFO & N/A & 1 & \\ 
\cite{Zeng-13-A} & MC & Freq. sel. & No & No & Pilot & Both & CFO & No & 1 & \\
\cite{Chen-12-Jan-A} & MC & Freq. sel. & No & No & Pilot & Both & Both & No & 1 & \\
\cite{Al-Habashna-12-A} & MC & Freq. sel. & No & No & N/A & Comp & Both & No & 1 & \\
\cite{Cheraghi-12-A} & MC & Freq. sel. & No & No & N/A & Comp & Both & No & 1 & \\
\cite{Ding-13-Dec-P} & MC & Freq. sel. & No & No & Pilot & Both & CFO & Yes & 1 & \\
\cite{Zhao-11-Dec-P} & MC & Freq. sel. & No & Yes & Pilot & Est & CFO & Yes & $>1$ & \\
\cite{Zivkovic-11-Dec-P} & MC & AWGN & N/A & N/A & Pilot & Est & CFO & No & 1 & TS design \\
\cite{Zhou-11-Dec-P} & MC & Freq. sel. & No & No & Pilot & Both & CFO & No & 1 & SDR implementation \\ 
\cite{Liu-12-May-P} & MC & Freq. sel. & No & No & Blind & Both & Both & No & 1 & \\
\hline
 \end{tabular}
\label{tab:CR}
\vspace{1cm}
\caption{Summary of synchronization research in Distributed multiuser communication systems} \centering
\begin{tabular}{|l|l|l|l|l|l|l|l|l|l|} \hline
Article & SC/MC & Fading  & CSI Req. & CE &  Blind/Pilot & Est/Comp & TO/CFO  & Bound & Comments  \\ \hline
\cite{Liu-12-Aug-A} & MC & Freq. sel. & Yes & No & N/A & Comp & CFO & N/A & TR-STBC \\
\cite{Caus-11-Jun-P} & MC & Freq. sel. & Yes & No & N/A & Comp & Both & N/A & FBMC SIMO \\
\cite{Zhi-11-May-P} & MC & Freq. sel. & No & Yes & Pilot & Both & Both & No & TD-LTE cell search \\
\cite{Sanchez-11-Jun-P} & MC & Freq. sel. & No & Yes & Pilot & Comp & Both & No & Cooperative STC \\
\hline
 \end{tabular}
\label{tab:DMU}
\vspace{1cm}
\caption{Summary of synchronization research in CoMP based communication systems} \centering
\begin{tabular}{|l|l|l|l|l|l|l|l|l|l|l|} \hline
Article & SC/MC & Fading  & CSI Req. & CE &  Blind/Pilot & Est/Comp & TO/CFO  & Bound & Type & Comments  \\ \hline
\cite{Zarikoff-10-A} & SC & Freq. flat & Yes & No & Pilot & Both & CFO & Yes & CB & downlink \\
\cite{Liu-13-Apr-P} & SC & Freq. flat & Yes & No & N/A & Comp & Both & No & JT & downlink, DLC-STC \\
\cite{Silva-14-A} & SC & Freq. sel. & Yes & Yes & Pilot & Both & TO & No & JT & downlink \\
\cite{Zhao-14-A} & MC & not discussed & Yes & No & N/A & Comp & TO & N/A & JT &  OFDMA downlink \\
\cite{Liang-12-A} & MC & Freq. sel. & No & Yes & Pilot & Both & CFO & Yes & CB &  MIMO-OFDM downlink \\
\cite{Iwelski-14-A} & MC & Freq. sel. & No & Yes & Pilot & Comp & TO & No & TPS & OFDM downlink \\
\cite{Jiang-14-A} & MC & Freq. sel. & No & Yes & semiblind & Both & CFO & No & CB & OFDM uplink \\
\cite{Tsai-13-A} & MC  & Freq. sel. & No & Yes & Pilot & Est & CFO & Yes & & OFDM downlink , TS design \\
\cite{Vakilian-13-Dec-P} & MC & Freq. sel. & Yes & No & N/A & Comp & CFO & N/A & & OFDM,UFMC uplink \\
\cite{Koivisto-13-Sep-P} & MC & Freq. sel. & No & Yes & Pilot & Both & Both & No & TPS & OFDM downlink \\
\cite{Pec-14-A} & MC & Freq. sel. & No & Yes & Pilot & Both & CFO & No & CB & OFDM downlink, DoA est. \\
\hline
 \end{tabular}
\label{tab:CoMP}
\vspace{1cm}
\caption{Summary of synchronization research in Multicell Interference based communication systems} \centering
\begin{tabular}{|l|l|l|l|l|l|l|l|l|l|l|} \hline
Article & SC/MC & Fading  & CSI Req. & CE &  Blind/Pilot & Est/Comp & TO/CFO  & Bound & Network & Comments  \\ \hline
\cite{Coelho-11-Sep-P} & SC & Freq. sel. & Yes & No & N/A & Comp & CFO & N/A & SC-FDE downlink &\\
\cite{Mochida-12-May-P} & SC & Freq. sel. & No & No & Pilot & Both & TO & No &  SC-FDMA uplink & \\
\cite{Deng-14-A} & MC & Freq. sel. & No & Yes & Pilot & Both & CFO & Yes &  OFDMA downlink & \\
\cite{Morelli-14-A} & MC & Freq. sel. & No & Yes & Pilot & Est & Both & Yes &  OFDMA downlink & \\
\cite{Hung-14-A} & MC & Freq. sel. & No & No & Pilot & Both & CFO & Yes &  OFDM downlink & TS design \\
\cite{Chang-12-Nov-A} & MC & Freq. sel. & No & No & Pilot & Comp & Both & No &  OFDM downlink & Cell search \\
\cite{Yang-13-A} & MC & Freq. sel. & No & Yes & Pilot & Both & CFO & No &  OFDM downlink & DoA Est \\
\cite{Liu-13-Apr-WCNC-P} & MC & Freq. sel. & No & No & Pilot & Comp & CFO & No & OFDM downlink & Cell Search \\ 
\hline
 \end{tabular}
\label{tab:MCI}
\end{table*}
\subsection{Multicell interference based communication systems}

In a multicell interference communication network, the received signal is contaminated by intercell interference. The interference can arise from the neighboring cells, when there is a universal frequency reuse, i.e., available frequency resources are reused in each cell in the network. This can cause interference amongst different tiers in a heterogeneous network, e.g., femtocell base stations may receive interfering signals from the macrocell user in addition to the desired signal from the femtocell user. The synchronization challenge for the receiver is to achieve timing and frequency synchronization with the desired signal in the presence of this interference. In addition to estimate and compensate TO and CFO between the desired user and the receiver, it also has to suppress intercell interference in order to decode the desired signal. To date most approaches to synchronization have modeled the interference as an additive term that can be combined with the noise. However, this approach may be very suboptimum for synchronization in heterogeneous networks.

The summary of the research carried out to achieve timing and carrier synchronization in multicell interference based communication systems is given in Table \ref{tab:MCI}:
\begin{enumerate}

\item Timing synchronization alone considering single carrier communication is analyzed in \cite{Mochida-12-May-P}.

\item Carrier synchronization alone is studied in \cite{Coelho-11-Sep-P,
Deng-14-A,Hung-14-A,Yang-13-A,Liu-13-Apr-WCNC-P}, where \cite{Coelho-11-Sep-P} consider single carrier communication and \cite{Deng-14-A,Hung-14-A,Yang-13-A,Liu-13-Apr-WCNC-P} consider multi-carrier communication.

\item Joint timing and carrier synchronization considering multi-carrier communication is analyzed in \cite{Morelli-14-A,Chang-12-Nov-A}.

\end{enumerate}

The information about the physical layer used, e.g., SC-FDE downlink, SC-FDMA uplink, or OFDMA/OFDM downlink is provided under ``Physical Layer" field in Table \ref{tab:MCI}. Finally, as detailed in Table \ref{tab:MCI}, the categorized papers may also differ in respect of providing joint channel estimation, requiring CSI or training, proposing estimation or compensation, or providing lower bounds on estimation of synchronization parameters. Further details about the consideration of TS design, cell search, or DoA estimation is provided in the last column of Table \ref{tab:MCI}.

\section{UWB and Non-CDMA Based Spread Spectrum Communication Systems}\label{sec:other}

\subsection{UWB communication systems}

\subsubsection{System Model} Ultra wide band (UWB) refers to a radio communication technique based on transmitting very short duration pulses, typically nano seconds or less, whereby the occupied bandwidth gets very large. The technology is used at a very low energy level for short-range, high-bandwidth communication using a large portion of the radio spectrum ($> 500$ MHz). UWB communication transmits in a manner that results in little to no interference to narrow band signals that may be operating in the same frequency band. The UWB technology allows communication either over baseband or RF. The baseband or carrierless communication system, known as impulse radio (IR) UWB communication system, employs time-hopping. The RF communication system mainly involves multi-carrier communication, referred to as multi-band (MB)-OFDM UWB communication system and usually employs frequency-hopping. The brief system model details for IR-UWB and MB-OFDM UWB communication systems are given as follows:

\paragraph{IR-UWB} In IR-UWB communications, a single data symbol is associated with several consecutive pulses, each located in its own frame. Accordingly, each data symbol is spread by sub nano second pulses. Further, spreading of these pulses is achieved by time hopping these low duty cycle pulses and data modulation is accomplished by additional pulse position modulation. Pulse width indicates the center frequency of the UWB signal. As IR based UWB does not use any carrier signal, it is also known as baseband, or carrierless or zero-carrier technology.

In a multipath environment, a fine resolution of multipath arrivals occurs due to large transmission bandwidth. This leads to reduced fading for each path because the transmitted data is in the form of pulses and significant overlap is prevented. Thus, to reduce the possibility of destructive combining~\cite{Roy-A-2004}. Rake receivers~\cite{Bottomley-2000} are employed to collect the signal energy of the multipath components, achieving much higher processing gain. Due to its significant bandwidth, an IR based multiple-access system may accommodate many users, even in multipath environments. Multiple access to the channel is made possible by changing the pulse position within a frame according to a user-specific time-hopping code.

\paragraph{MB-OFDM UWB} Multiband OFDM (MB-OFDM) approach for UWB is based on multiple OFDM bands each with at least $500$ MHz bandwidth and each OFDM band comprising multiple sub-carriers. It can also be thought of as a combination of frequency hopping with the sub-carriers occupying one band at one time and hopping according to a pre-defined hopping pattern.

\subsubsection{Synchronization Challenge}

The synchronization challenge of the two different UWB technologies, IR-UWB and MB-OFDM-UWB, is given in the following two subsections.

\paragraph{IR-UWB} Timing errors as small as a fraction of a nanosecond can seriously degrade the system performance. Timing recovery can be viewed as a two-part process. The first part consists of estimating the beginning of the individual frames relative to the receiver's clock ticks. This is called frame timing. The second part consists of identifying the first symbol of each frame in the incoming frame stream and is referred to as symbol timing.

Frequency offset in IR-UWB communication system arises due to the clocks at the transmitter and receiver that run independently at slightly different frequencies, albeit close to a common nominal value.  Deviations from the nominal value are referred to as clock frequency offsets. Moreover, Doppler fading also plays a key role in causing frequency offset in IR-UWB communication systems.

\paragraph{MB-OFDM UWB} The MB-OFDM systems have the following distinctive characteristics compared to conventional OFDM systems: 1) different channel responses and channel energies across the different bands, 2) different carrier frequency offsets across the different bands, and 3) the interplay between timing and frequency hopping (a mismatched timing point at the receiver will yield mismatched frequency hopping and hence a significant performance degradation). These characteristics provide diversity but also additional design constraints, and hence, should be taken into account in the designs of synchronization, channel estimation, and equalization. Further, MB-OFDM based UWB receivers are quite sensitive to carrier frequency estimation errors \cite{Li-08-Nov-A}.


\subsubsection{Literature Review} The summary of the research carried out to achieve timing and carrier synchronization in UWB communication systems is given in Table \ref{tab:UWB}:
\begin{enumerate}

\item Timing synchronization alone considering IR-UWB communication system is studied in \cite{Chen-10-May-A,Lv-11-A}.

\item Carrier synchronization alone is studied in \cite{Erseghe-11-A,
Erseghe-11-D-A,Oh-10-May-A,You-11-A,Kim-11-J-A,Lin-12-A,Karim-10-A,Wang-10-A,Hwang-12-A}, where as identified in second column of Table \ref{tab:UWB}, \cite{Erseghe-11-A,Erseghe-11-D-A,Oh-10-May-A} consider IR-UWB communication and \cite{You-11-A,Kim-11-J-A,Lin-12-A,Karim-10-A,Wang-10-A,Hwang-12-A} consider MB-OFDM UWB communication.

\item Joint timing and carrier synchronization considering IR-UWB and MB-OFDM UWB communication is proposed in \cite{DAmico-13-A} and \cite{Ye-10-A,Fan-12-A}, respectively.

\end{enumerate}

As detailed in Table \ref{tab:UWB}, the categorized papers differ in respects of providing joint channel estimation, requiring CSI or training, proposing estimation or compensation, or providing lower bounds. If applicable, further details, e.g., consideration of MAI, TWR protocols, sampling frequency offset (SFO) estimation, or hardware implementations, are provided in the last column of Table \ref{tab:UWB}. Note that when considering the TWR protocol, range estimation is composed of both time of arrival (TOA) estimation of a direct path as well as round trip time (RTT) estimation.

\subsection{Non-CDMA Based Spread Spectrum Communication Systems}  In this section, the literature survey focuses on spread spectrum communication systems aside from CDMA. Note that the list of papers dealing with synchronization in CDMA systems is already provided in Section \ref{sec:CDMA}.

Spread spectrum techniques can be applied to overcome a jamming situation, i.e., when an adversary intends to disrupt the communication. There, the aim of spread spectrum communication is to make the transmitted signal such that it should be difficult to detect by an adversary, i.e., the signal should have a low probability of interception and should be difficult to jam. The term \emph{spread spectrum} stems from the fact that the transmitted signal occupies a much wider frequency band than what is necessary. This enables the transmitter to hide its signal in a larger bandwidth. CDMA is realised by spread spectrum techniques, but security aspects and anti­jamming/anti-noise properties of SS communication may have become less important to CDMA communications since it has been adopted as a mobile standard.

Different spread spectrum techniques that use time-domain, frequency-domain, direct pseudonoise chip sequence, and multi-carrier have been proposed. Among the above techniques, frequency-domain spread spectrum employ multi-carrier communication. Similar to CDMA communication systems, non-CDMA based spread-spectrum based communication is also prone to synchronization errors. Thus, the synchronization challenge of non-CDMA based spread spectrum communication is to estimate and compensate the effect of TO and CFO from the received signal before despreading it. Non-CDMA based spread spectrum systems may feature their specific synchronization methods, e.g., a specific frequency synchronization technique that can be applied only to non-CDMA based frequency-domain spread spectrum systems is presented in \cite{Kohda-13-Sep-P}.

Table \ref{tab:SS} summarizes the research carried out to achieve synchronization in non-CDMA based spread spectrum communication systems:
\begin{enumerate}

\item The timing synchronization alone considering single carrier communication is studied in \cite{Benedetto-11-A}.

\item The carrier synchronization alone considering single and multi-carrier communication is proposed in \cite{Oh-12-A-A} and \cite{Jang-13-A}, respectively.

\item  The joint timing and carrier synchronization considering single and multi-carrier communication is proposed in \cite{Kohda-13-Sep-P} and \cite{Kohd-12-Dec-P}, respectively.

\end{enumerate}

As detailed in Table \ref{tab:SS}, the categorized papers differ in respect of proposing estimation or compensation design. If applicable, further details whether Gabor division spread spectrum system (GD-S$^3$) or frequency division spread spectrum system (FD-S$^3$) is considered, are provided in last column of Table \ref{tab:SS}.


\begin{table*}[t]
\vspace{1cm}
\caption{Summary of synchronization research in UWB communication systems} \centering
\begin{tabular}{|l|l|l|l|l|l|l|l|l|l|} \hline
Article & IR/MC & Fading & CSI Req. & CE &  Blind/Pilot & Est/Comp & TO/CFO  & Bound &  Comments  \\ \hline
\cite{Erseghe-11-A} & IR  & Freq. sel. & No & Yes & Pilot & Both & CFO & Yes &  MAI \\ 
\cite{DAmico-13-A} & IR  & Freq. sel. & No & No & Pilot & Est & Both & No  &  \\
\cite{Erseghe-11-D-A} & IR  & Freq. sel. & No & Yes & Pilot & Est & CFO & Yes &  MAI \\ 
\cite{Chen-10-May-A} & IR  & Freq. sel. & Yes & No & Pilot & Both & TO & No   & \\
\cite{Oh-10-May-A} & IR  & not discussed & N/A & N/A & Pilot & Est & CFO & No &  TWR protocol \\
\cite{Lv-11-A} & IR  & Freq. sel. & No & No & Pilot & Both & TO & No & \\ 
\cite{You-11-A}  & MC & Freq. sel. & No & No & Pilot & Both & CFO & No & SFO   \\
\cite{Kim-11-J-A}  & MC & Freq. sel. & No & Yes & Pilot & Both & CFO & No &  Blind CFO tracking \\
\cite{Lin-12-A}  & MC & Freq. sel. & No & Yes & Pilot & Both & CFO & Yes &  SFO   \\
\cite{Karim-10-A}  & MC & Freq. sel. & Yes & No & Pilot & Est & CFO & No &    \\
\cite{Wang-10-A}  & MC & Freq. sel. & No & Yes & Pilot & Both & CFO & No & SFO   \\
\cite{Hwang-12-A} & MC & Freq. sel. & No & Yes & Pilot & Both & CFO & No & SFO, Hardware implementation  \\
\cite{Ye-10-A} & MC & Freq. sel. & No & Yes & Pilot & Both & Both & No &   \\
\cite{Fan-12-A} & MC & Freq. sel. & No & Yes & Pilot & Both & Both & No &  Hardware implementation  \\
\hline
 \end{tabular}
\label{tab:UWB}
\vspace{1cm}
\caption{Summary of synchronization research in spread spectrum communication systems} \centering
\begin{tabular}{|l|l|l|l|l|l|l|l|l|l|} \hline
Article & SC/MC & Fading & CSI Req. & CE &  Blind/Pilot & Est/Comp & TO/CFO  & Bound & Comments  \\ \hline
\cite{Oh-12-A-A} & SC & Freq. sel. & Yes & No & Pilot & Comp & CFO & N/A & TWR method \\
\cite{Benedetto-11-A} & SC & not discussed & N/A & N/A & Pilot & Est & TO & No & \\
\cite{Kohda-13-Sep-P} & SC & Freq. flat & Yes & No & Pilot & Est & Both & No & GD-S$^3$  \\
\cite{Kohd-12-Dec-P} & MC & Freq. sel. & Yes & No & Pilot & Comp & Both & No & FD-S$^3$  \\
\cite{Jang-13-A} & MC & Freq. sel. & Yes & No & N/A & Comp & CFO & N/A & FD-S$^3$  \\
\hline
 \end{tabular}
\label{tab:SS}
\vspace{1cm}
\caption{Summary of synchronization research in $60$ GHz communication systems} \centering
\begin{tabular}{|l|l|l|l|l|l|l|l|l|l|l|} \hline
Article & SC/MC & Fading & CSI Req. & CE &  Blind/Pilot & Est/Comp & TO/CFO  & Bound & System Model & Comments  \\ \hline
\cite{Koschel-12-Sep-P} & MC & Freq. sel. & No & Yes & Pilot & Both & CFO & No & MIMO-OFDM & PHN  \\
\cite{Ban-13-Jun-P} & MC & Freq. sel. & No & Yes & Pilot & Both & CFO & No & OFDM & SFO, hardware implementation  \\
\hline
 \end{tabular}
\label{tab:60G}
\end{table*}

\section{Future Directions}\label{sec:future}

With the further adoption of wireless technologies in new and emerging fields, it is anticipated that cellular systems may need to operate in a more distributed fashion. Moreover, to support a growing number of devices, cellular systems are expected to take advantage of new frequency bands and technologies. Below we discuss some promising directions for future research.

\textbf{Millimeter-Wave and Terahertz Frequencies}: The shortage of bandwidth has encouraged new research in the field of millimeter-wave and terahertz communications. Unlike the microwave band, there is a very large amount of under utilized bandwidth at these frequencies. However, millimeter-wave ($30$ GHz to $300$ GHz) and terahertz ($300$ GHz to $1$ THz) communication systems are affected by significant signal attenuation due to path loss and shadowing~\cite{Mehrpouyan-14-Mar-Art}. Moreover, oscillators at such high carrier frequencies are not as accurate compared to their counterparts at microwave frequencies and, as such, millimeter-wave and terahertz are significantly more impacted by CFO and phase noise~\cite{Mehrpouyan-14-Mar-Art}. In addition to all of the above, amplifiers at millimeter-wave and terahertz are expected to operate in the saturation region to generate enough power to overcome the shadowing and path loss issues in these bands. As such, the received signal at these frequencies may be severely affected by nonlinearities~\cite{Rappaport-2014-B}. Hence, new synchronization algorithms for this band are needed that can operate at very low SNR, can still achieve end-to-end synchronization in the presence of severe amplifier nonlinearity, and can track impairments such as carrier frequency offset and timing offset in the presence of significant phase noise.

As illustrated in Table \ref{tab:60G}, the research in synchronization for mmwave systems is in an early stage. Both listed papers \cite{Koschel-12-Sep-P,Ban-13-Jun-P} consider multi-carrier communication and propose joint channel estimation with CFO estimation and compensation design. In particular, \cite{Koschel-12-Sep-P} considers MIMO system with the presence of PHN, while \cite{Ban-13-Jun-P} also proposes sampling frequency offset estimation and considers hardware implementation.

\textbf{Massive MIMO}: One approach to achieve higher bandwidth efficiency and to support more users per base station is to use a very large number of antennas at the base station. Theoretical studies and measurement campaigns have shown that massive MIMO systems can overcome significant interference and the additive noise in cellular networks \cite{Marzetta-10-Nov-Art}. However, massive MIMO systems require the estimation of a very large number of channel and synchronization parameters at the cost of a significant overhead. Due to the large number of antennas and users in a massive MIMO system, it is not possible to allocate distinct and orthogonal training sequences to each user. This creates pilot overlap, so massive MIMO systems are impacted by pilot contamination. 

New algorithms are, therefore, needed to overcome pilot contamination, complexity and overhead associated with synchronization in massive MIMO systems. The main approach to date has been on using time division duplex (TDD) and the resulting channel reciprocity~\cite{Marzetta-10-Nov-Art}. However, it is expected that as communication systems migrate to millimeter-wave and terahertz frequencies, frequency division duplex (FDD) systems may be deployed more often than TDD systems. In addition, most of the synchronization work to date has been based on the assumption that the massive MIMO channel is full-rank, which is contrary to practical measurement results. In fact, massive MIMO channels are expected to be severely sparse \cite{Torkildson-10-Nov-Art}, and channel sparsity ought to be used to reduce synchronization overhead.

\textbf{Full-Duplex Communications}: Although the next generation cellular networks are expected to migrate to the millimeter-wave and terahertz bands, the microwave band will continue to play an important role in 5G and beyond due to its favorable propagation characteristics. Hence, to achieve better bandwidth efficiency at microwave frequencies, full-duplex communication systems are considered for broad adoption in next generation cellular systems \cite{5G-May-14}. Although significant research has been carried out~\cite{Sabharwal-2014,Liu-2015S}, the issue of synchronization in full-duplex communication systems continues to be an open area of research. This can be attributed to the challenges associated with estimating synchronization parameters in the presence of severe self-interference encountered in full duplex communication systems. One solution could be to leverage the advancements in the field of antenna design and use reconfigurable and directional antennas at the transceiver to reduce self-interference. This combined with algorithms that can blindly track the synchronization parameters in the presence of self-interference may advance the research in this field.

\textbf{Radio Frequency Energy Harvesting Communication Systems}: Radio frequency (RF)-enabled simultaneous information and power transfer (SWIPT) has emerged as an attractive solution to power nodes in future wireless networks~\cite{Lu-2015S,AliNasir-2013,ZhangRui-2013}. The design of synchronization schemes for SWIPT has not yet been considered in the literature. This is a challenging task since information and power transfer may have fundamentally different objectives, e.g., interference degrades the quality of the information transfer but can actually be beneficial from the viewpoint of RF energy harvesting. The speed and reliability of the information transfer obviously depends on accurate synchronization. One might (misleadingly) think that synchronization may not be necessary for power transfer since it does not affect the process of RF-DC conversion, e.g., a time-misaligned sine wave can still be rectified. However, carrier mismatch may mean that a receiver is not able to lock onto the RF signal in the first place. Thus, accurate synchronization is clearly needed for \textit{efficient} power transfer. In this regard, fundamental problems include how to determine the optimum energy harvesting, pilot training, and data transmission times, in order to achieve maximum throughput and high synchronization accuracy in wireless energy harvesting systems.


\textbf{Vehicular Communications}: Vehicular communications is key to the development of smart roads and autonomous vehicles and, as such, will be a key service that will be provided by next generation cellular networks. Enabling vehicular communication requires synchronization in highly mobile environments with very low latency. There are very few synchronization algorithms that can meet both needs simultaneously and significant research in this field is needed \cite{Zheng-15-EA-Art}.

\textbf{Cognitive Radio Networks}: Cognitive radio is a key technology in enhancing bandwidth utilization. A variety of spectrum allocation and detection algorithms have been proposed for interweave cognitive networks, which allow secondary users to sense the environment and transmit only when the primary users are silent. However, as identified in Section~\ref{sec:CR}, the majority of these approaches are based on the assumption of perfect synchronization. This assumption must be relaxed to realize more practical spectrum detection algorithms for cognitive radio networks~\cite{Wu-15-Art}. In addition, synchronization aspects for underlay and overlay cognitive networks need to be explored.

\textbf{Satellite Systems}: Although today's satellite networks are mainly used for broadcasting applications, their applications are expected to expand to provide two way cellular connections in next generation networks. The main advantage of satellite systems is their ability to provide wide-area coverage in remote areas without the need to deploy base stations. However, the popularity of satellite systems has resulted in adjacent satellite interference \cite{Shaban-14-Conf}. This interference is much more difficult to combat since the number of interferers, i.e., satellites, exceeds the number of receiver antennas (generally one dish antenna). This results in an overloaded receiver scenario, which is generally different from the case of cellular networks. Moreover, satellite systems may not straightforwardly apply sophisticated physical layer techniques such as OFDM due to significant peak-to-average power ratio (PAPR). As such, synchronization algorithms that can enable locking into the desired satellites signal in the presence of interference are highly desirable.

\section{Conclusions}\label{sec:con}

In this survey, we have provided a comprehensive classification of the timing and carrier synchronization research published in the literature in the last five years. Important contributions of this work include Tables II-XXI, which summarise the system model assumptions and the state-of-the-art synchronization solutions and their limitations for the different communication systems. In addition, directions for future research have been outlined.

\ifCLASSOPTIONpeerreview
\else
\balance
\fi

\bibliographystyle{IEEEtran}


\end{document}